\documentclass[11pt,a4paper]{article}
\pdfoutput=1
\usepackage{jheppub}
\usepackage{amsmath}
\usepackage{graphicx}
\usepackage{subcaption}
\usepackage[toc,page]{appendix}
\usepackage{empheq}
\newcommand*\widefbox[1]{\fbox{\hspace{1em}#1\hspace{1em}}}

\allowdisplaybreaks

\title{Vacuum Stability and Symmetry Breaking in Left-Right Symmetric Model }

\author[a]{Garv Chauhan}
\affiliation[a]{Department of Physics and McDonnell Center for the Space Sciences,  Washington University, St. Louis, MO 63130, USA}
\emailAdd{garv.chauhan@wustl.edu}

\date{\today}

\begin{document}

\abstract{We derive analytic necessary and sufficient conditions for the vacuum stability of the left-right symmetric model by using the concepts of copositivity and gauge orbit spaces. We also derive the conditions sufficient for successful symmetry breaking and the existence of a correct vacuum. We then compare results obtained from the derived conditions with those from numerical minimization of the scalar potential. Finally, we discuss the renormalization group analysis of the scalar quartic couplings through an example study that satisfies vacuum stability, perturbativity, unitarity  and experimental bounds on the physical scalar masses.}

\maketitle

\section{Introduction}
The Standard Model (SM) has been one of the most successful theory with its predictions in remarkable agreement with the experimental data. Yet the SM leaves many open questions to be answered. Discovery of neutrino oscillations has decisively proved the existence of neutrino masses\footnote{The lightest neutrino may still be massless.}. This is in glaring contradiction with SM which only features massless neutrinos. 
\par Left-Right Symmetric Model (LRSM) is the simplest extension of the SM with modified electroweak gauge group: $ SU(2)_L \otimes SU(2)_{R} \otimes U(1)_{B-L}$ \cite{Pati:1974yy,Mohapatra:1974gc,Senjanovic:1975rk,Davidson:1978pm}. It features heavy Majorana right-handed neutrinos and can naturally explain the small masses of left-handed neutrinos through see-saw mechanism \cite{Minkowski:1977sc,Mohapatra:1979ia,Yanagida:1979as,GellMann:1980vs,Glashow:1979nm}. It explains the asymmetric chiral structure of SM through restoration of parity symmetry at high energies. 
\par An important problem with the SM is the stability of the scalar Higgs potential at high-energies. The condition for stability of the scalar
potential in the SM is the positivity of the Higgs quartic coupling $\lambda_h$. However, renormalization group equation (RGE) analysis shows that $\lambda_h$ becomes negative at a scale of around $10^{10}$ GeV for experimentally measured value of the Higgs mass  \cite{Isidori:2001bm}. Thus, the potential in the SM is unbounded from below around this scale and makes the theory unstable. This motivates us to to ensure the stability of the scalar Higgs potential in LRSM as a candidate theory for physics beyond the SM.
\par  Scalar sector of LRSM features an SU(2) bi-doublet, left and right-handed weak isospin triplets. Such an extended scalar sector leads to a complicated form of the potential which contains 17 free parameters (3 negative mass squares and 14 scalar quartic couplings). Analytical study of vacuum stability and desired minimum for the entire scalar potential is an arduous 
task. There has been some work in this direction \cite{Chakrabortty:2013mha,Dev:2018foq} but the results only hold for a small parameter space with most of the quartic couplings set to zero. Moreover, just ensuring vacuum stability does not yield the desirable vacuum expectation values (VEVs) to ensure correct spontaneous symmetry breaking to SM \cite{Dev:2018foq}. In this work, we have derived most general conditions sufficient\footnote{We have set only few of the couplings($\alpha_2,\beta_i$'s) to zero.} to obtain the correct symmetry breaking and ensure vacuum stability of the LRSM. As we show later, it is necessary to obtain conditions for vacuum stability of the general scalar potential before requiring the correct VEV alignment at the minimum. The procedure outlined here for finding conditions for correct symmetry breaking is general in nature and can be applied to different theories with varied forms of the scalar sector.
\par This work is organised as follows. In section \ref{bound}, concepts of copositivity and gauge orbit spaces are presented in context of vacuum stability. In section \ref{LR}, we review the model details of LRSM. In section \ref{VS}, we derive the necessary and sufficient conditions for the boundedness of scalar potential of the LRSM. In section \ref{SB}, we derive conditions sufficient for scalar parameters to lead to spontaneous symmetry breaking (SSB) to the correct global minimum. In section \ref{Num}, we compare the results from numerical minimization of the potential with those from the derived conditions. In section \ref{RGE}, we present an example study to use these conditions and other theoretical constraints (unitarity, scalar mass spectrum, perturbativity) on the quartic couplings to study the stability of the vacuum at high energies and agreement with current experimental limits on scalar mass spectrum. Finally, we conclude in section \ref{end}.

\section{Boundedness}\label{bound}
For the stability of the vacuum state, the potential should be bounded in all field directions. In the large-field limit, terms with dimension $d<4$ can be ignored as they are negligible in comparison to the quartic terms ( denoted by $V_4(\phi_i)$) in the potential. Thus, requiring $V_4(\phi_i)>0$ as field values $\phi_i \rightarrow \infty$ is a strong condition for boundedness. This criterion is termed as Bounded From Below (BFB) condition.
\par For obtaining conditions for vacuum stability of a scalar potential using BFB criterion, concepts of copositivity criteria and gauge orbit spaces can help greatly simplify the analysis. 

\subsection{Copositivity Criteria}

Given a condition of the form: 
\begin{equation}\label{quad}
a x^2 + b x + c>0    
\end{equation}
where $x \in \mathbb{R}$, the conditions for it to be positive-definite are very well known. If $x \in \mathbb{R}^+$, then the requirement that eq. \eqref{quad} holds is termed as copositivity. The conditions for copositivity are given below:

$$ a>0,\, c>0,\,b + 2\sqrt{ac} >0  $$

The quartic part of the vacuum potential is bounded from below if it satisfies the copositivity conditions. The criteria of copositivity has been applied to numerous models in literature to obtain vacuum stability conditions \cite{Kannike:2012pe,Chakrabortty:2013mha,Kannike:2016fmd,Sanchez-Vega:2018qje}. The difficulty to solve these conditions based solely on copositive criteria is a formidable task. Usually it involves checking copositivity in all n-field directions to obtain an exhaustive list of conditions for vacuum stability . 
\par In sec. \ref{subL} and \ref{subR}, copositive criteria is used in conjunction with suitable parametrization of gauge orbit parameters to yield results easily. In the coupled case (Sec. \ref{dread}), when mixed field terms are present\footnote{i.e $\alpha$'s $\neq0$}, we observe that exact values of minima are required and copositivity isn't helpful as it yields results only upto a multiplicative constant.

\subsection{Gauge Orbit Spaces}\label{GOS}

Due to the gauge freedom of the theory, different values of the fields can lead to same value of the potential. These field values connected through gauge transformations collectively form a gauge orbit. Minimization of the Higgs potentials in orbit spaces has been extensively studied in context of grand unified theories in the 1980's \cite{Kim:1981xu,Kim:1981jj,Frautschi:1981jh,Kim:1983mc,Abud:1981tf,Abud:1983id}. Here, we present the method of orbit spaces for the two higgs fields case  \cite{Kim:1981xu}. This is an extension of the one-field treatment as presented in ref.\cite{Kim:1981xu,Kannike:2016fmd}. 
\par Consider the scalar potential of a theory with two higgs fields $\phi$ and $\pi$ charged under non-abelian gauge groups $G$ and $G^\prime$ respectively, with the following form :
\begin{eqnarray}\label{ghigg}
 V(\phi,\pi) & \ = \ & -\mu_1^2(\phi_i^*\phi_i) -\mu_2^2(\pi_i^*\pi_i) + \lambda_1(\phi_i^*\phi_i)^2 + \lambda_2 f_{ijkl}\phi_i^*\phi_j\phi_k^*\phi_l + \cdots  \\ \nonumber
 &  & + \rho_1(\pi_i^*\pi_i)^2 + \rho_2 g_{ijkl}\pi_i^*\pi_j\pi_k^*\pi_l  + \cdots   \\ \nonumber
 &  & + \alpha_1(\phi_i^*\phi_i)(\pi_j^*\pi_j)+ \cdots (\text{other terms coupling $(\phi,\pi)$})
\end{eqnarray}
where $V(\phi,\pi)$ remains invariant under the action of the group elements of $G$ and $G^\prime$. Field $\phi$($\pi$) (with components denoted by $\phi_i$($\pi_i$)) live in the representation \textbf{R}(\textbf{R$^\prime$}) of group $G$($G^\prime$).

\par The group elements of $G$ rotate a field into other field values on the same orbit space. It can be shown that all the fields $\psi_i$ on the orbit respect the same group, called the little group. If their action on the fields is unitary, the norm of the field value $\phi_i^*\phi_i$ is preserved. This similarly holds for field $\pi$. Several different orbits respect the same group and form a set. The set of these orbits is called the stratum of the little group. Thus, we need to find the gauge orbit that minimizes the potential. \newline
The dimensionless ratios of invariants called orbit space parameters specifies a strata as follows:
$$ A_n(\hat{\phi})= \frac{f_{ijkl}\phi_i^*\phi_j\phi_k^*\phi_l}{(\phi_i^*\phi_i)^2}
\quad B_n(\hat{\pi})= \frac{g_{ijkl}\pi_i^*\pi_j\pi_k^*\pi_l}{(\pi_j^*\pi_j)^2}
$$
Similarly, for coupled terms $C_n(\hat{\phi},\hat{\pi})$ can be defined but normalized by $\phi_i^*\phi_i\pi_j^*\pi_j$. Orbit space parameters greatly reduce the number of parameters and contain all the directional information required for the minimization. Defining orbit space parameters for eq. \eqref{ghigg},
\begin{eqnarray}\label{osV}
V(\phi,\pi) & \ = \ & -\mu_1^2|\phi|^2 -\mu_2^2|\pi|^2 +  |\phi|^4(\lambda_1 + \lambda_2 A_1(\hat{\phi}) + \lambda_3 A_2(\hat{\phi}) + \cdots) \nonumber \\ \nonumber 
&  & + |\pi|^4(\rho_1 + \rho_2 B_1(\hat{\pi}) + \rho_3 B_2(\hat{\pi}) + \cdots) \\ \nonumber
&  & + |\phi|^2|\pi|^2(\alpha_1 + \alpha_2 C_1(\hat{\phi},\hat{\pi})+ \cdots) \\ \nonumber \\
& \equiv & -\mu_1^2|\phi|^2 -\mu_2^2|\pi|^2 +  |\phi|^4 A(\lambda,\hat{\phi})+|\pi|^4 B(\rho,\hat{\pi}) + |\phi|^2|\pi|^2 C(\alpha,\hat{\phi},\hat{\pi}) 
\end{eqnarray}
where 
$$ |\phi|^2=\phi_i^*\phi_i,\quad |\pi|^2=\pi_i^*\pi_i, \quad \hat{\phi}= \frac{\phi}{|\phi|},\quad \hat{\pi}= \frac{\pi}{|\pi|} $$
$$A(\lambda,\hat{\phi}) = \lambda_1 + \lambda_2 A_1(\hat{\phi}) + \lambda_3 A_2(\hat{\phi}) + \cdots$$
$$B(\rho,\hat{\pi}) =  \rho_1 + \rho_2 B_1(\hat{\pi}) + \rho_3 B_2(\hat{\pi}) + \cdots$$
$$ C(\alpha,\hat{\phi},\hat{\pi}) = \alpha_1 + \alpha_2 C_1(\hat{\phi},\hat{\pi})+ \cdots  $$
Note that we have assumed terms like $|\phi|^3|\pi|$  and $|\phi||\pi|^3$ to be absent from the expression for $V(\phi,\pi)$. This is particularly true if the higgs potential is invariant under a reflection symmetry for $\phi$ and $\pi$. Requiring boundedness and applying copositivity criterion, we get the following conditions for the stability of the potential,
$$ |\phi|^4 A(\lambda,\hat{\phi})+|\pi|^4 B(\rho,\hat{\pi}) + |\phi|^2|\pi|^2 C(\alpha,\hat{\phi},\hat{\pi}) >0 \quad \forall \, A(\lambda,\hat{\phi}),B(\rho,\hat{\pi}),C(\alpha,\hat{\phi},\hat{\pi})$$
\begin{equation}\label{cpV}
\implies A>0,\, B>0,\,C + 2\sqrt{AB
} >0    
\end{equation}
Treatment in ref.\cite{Kim:1981xu} assumes the monotonicity of the orbit space parameters in the potential and thus minimization of these parameters are not required. Our treatment for the left-right model differs here due to the presence of non-linearity in orbit space parameters. It should be noted that eq. \eqref{cpV} must also be minimized over all orbit space parameters. We also study the VEV structure of the scalar fields in the theory. Thus, minimizing $V$ w.r.t to $|\phi|$ and $|\pi|$ yields,
$$ \frac{\partial{V}}{\partial{|\phi|}}=2|\phi|\left(-\mu_1^2+ 2|\phi|^2 A+ |\pi|^2 C\right)=0$$
$$ \frac{\partial{V}}{\partial{|\pi|}}=2|\pi|\left(-\mu_2^2+ 2|\pi|^2 B+ |\phi|^2 C\right)=0$$
Since, field value should be non-zero, the minimum occurs at:
\begin{equation}\label{fmin}
|\phi_0|^2=\frac{2B \mu_1^2- C\mu_2^2}{4AB-C^2} \quad
|\pi_0|^2=\frac{2A \mu_2^2- C\mu_1^2}{4AB-C^2}
\end{equation}
Using second derivative analysis for $\phi$ and $\pi$, it can be proved that field values in eq. \eqref{fmin} leads to a minimum of the potential if and only if following conditions are satisfied.
\begin{equation}\label{muc1}
2B \mu_1^2- C\mu_2^2> 0   
\end{equation}
\begin{equation}\label{muc2}
2A \mu_2^2- C\mu_1^2 > 0
\end{equation}
\begin{equation}\label{muc3}
4AB-C^2 >0
\end{equation}
Plugging obtained field values at the minimum in eq. \eqref{osV}, we get
\begin{equation}\label{eq:min}
    V_0(\phi) = -\frac{B\mu_1^4-C\mu_1^2\mu_2^2+A\mu_2^4}{4AB-C^2}
\end{equation}
It can be shown using conditions obtained above that this minimum is guaranteed to be the global minimum of the potential.

\section{Left-Right Symmetric Model}\label{LR}

Left-Right Symmetric model (LRSM) is a gauge extension of the Standard Model (SM), which restores parity symmetry at high-energies \cite{Pati:1974yy,Mohapatra:1974gc,Senjanovic:1975rk}. It treats left and right handed chiralities of fermions equally prior to spontaneous symmetry breaking. It features heavy right-handed Majorana neutrinos, and thus explains small masses of left-handed neutrinos via the
see-saw mechanism \cite{Minkowski:1977sc,Mohapatra:1979ia,Yanagida:1979as}. The extended gauge group for this model : $SU(3)_C \otimes SU(2)_L \otimes SU(2)_{R} \otimes U(1)_{B-L}$. The particle content and their irreducible representations under the gauge group is given in table \ref{tab:parC}. The spontaneous symmetry breaking (SSB) of LRSM proceeds in two steps. First, the electrically neutral component of $\Delta_R$ acquires a VEV $v_R$ and breaks the gauge group from $SU(2)_{R} \otimes U(1)_{B-L}$ to $U(1)_{Y}$. Finally, the VEV of bidoublet $\Phi$ breaks the symmetry down to $U(1)_{Q}$ \cite{Gunion:1989in,Deshpande:1990ip}. The VEV structure of the scalar fields is 
\begin{equation}
\Phi \ = \ \frac{1}{\sqrt{2}} \left(\begin{array}{cc}
\kappa_{1} & 0\\
0 & \kappa_{2}e^{i\theta_{2}}
\end{array}\right),\qquad \Delta_{L} \ = \ \frac{1}{\sqrt{2}} \left(\begin{array}{cc}
0 & 0\\
v_{L}e^{i\theta_L} & 0
\end{array}\right),\qquad \Delta_{R} \ = \ \frac{1}{\sqrt{2}} \left(\begin{array}{cc}
0 & 0\\
v_{R} & 0
\end{array}\right)\label{eq:LRV-8}
\end{equation}
\newline
Note that only the neutral components acquire VEV so that $U(1)_{\text{EM}}$ does not break. Using the gauge transformations, two of the phases in $\kappa_1$ and $v_R$ have been rotated away. It is required that the VEV's respect the following hierarchy for correct phenomenology: $$v_L\ll\kappa_{1,2}\ll v_R$$
The electric charge formula takes the form:
$$ Q = T_{3L}+T_{3R}+\frac{B-L}{2}$$
where $T_{3X},\:X=(L,R)$ is the third generator of the group $SU(2)_X$ and $B-L$ is the baryon minus lepton number, the charge for group $U(1)_{B-L}$ \cite{Marshak:1979fm,Mohapatra:1980qe}.
\begin{table}[!t]
 \centering
  \begin{tabular}{l c c c c}
  \hline\hline
  &  $SU(3)_C$ &  $SU(2)_L$ & $SU(2)_{R}$ & $U(1)_{B-L}$\\ \hline
  $Q_L \equiv \begin{pmatrix} u_L \\ d_L \end{pmatrix}$ & $\mathbf{3}$ & $\mathbf{2}$ & $\mathbf{1}$ & $\frac13$ \\
  $Q_R \equiv \begin{pmatrix} u_R \\ d_R \end{pmatrix}$  & $\mathbf{3}$ & $\mathbf{1}$ & $\mathbf{2}$ & $\frac13$ \\ \hline
  $\psi_L \equiv \begin{pmatrix} \nu_L \\ e_L \end{pmatrix}$ & $\mathbf{1}$ & $\mathbf{2}$ &$\mathbf{1}$ & $-1$ \\
  $\psi_R \equiv \begin{pmatrix} N \\ e_R \end{pmatrix}$ & $\mathbf{1}$ & $\mathbf{1}$ &$\mathbf{2}$ & $-1$ \\  \hline
  $\Phi = \left(\begin{matrix}\phi^0_1 & \phi^+_2\\\phi^-_1 & \phi^0_2\end{matrix}\right)$ & $\mathbf{1}$ & $\mathbf{2}$ & $\mathbf{2}$ & 0 \\
  $\Delta_{L} = \left(\begin{matrix} \frac{1}{\sqrt2} \Delta^+_{L} & \Delta^{++}_{L} \\ \Delta^0_{L} & - \frac{1}{\sqrt2} \Delta^+_{L} \end{matrix}\right)$  & $\mathbf{1}$ & $\mathbf{3}$ & $\mathbf{1}$ & 2 \\ 
   $\Delta_{R} = \left(\begin{matrix} \frac{1}{\sqrt2} \Delta^+_{R} & \Delta^{++}_{R} \\ \Delta^0_{R} & - \frac{1}{\sqrt2} \Delta^+_{R} \end{matrix}\right)$  & $\mathbf{1}$ & $\mathbf{1}$ & $\mathbf{3}$ & 2 \\ \hline
  \end{tabular}
  \caption{Particle content of left-right symmetric model based on the gauge group $SU(3)_C \otimes SU(2)_L \otimes SU(2)_{R} \otimes U(1)_{B-L}$. }
  \label{tab:parC}
\end{table}
The most general renormalizable scalar potential for LRSM contains 17 independent terms \cite{Deshpande:1990ip,Dev:2018foq}: 
 
\begin{eqnarray}\label{eq:LRV-4}
V & \ = \  & -\mu_{1}^{2}\text{Tr}[\Phi^{\dagger}\Phi]-\mu_{2}^{2}\left(\text{Tr}[\tilde{\Phi}\Phi^{\dagger}]+\text{Tr}[\tilde{\Phi}^{\dagger}\Phi]\right)-\mu_{3}^{2}\left(\text{Tr}[\Delta_{L}\Delta_{L}^{\dagger}]+\text{Tr}[\Delta_{R}\Delta_{R}^{\dagger}]\right)+\lambda_{1}\text{Tr}[\Phi^{\dagger}\Phi]^{2}\nonumber \\
 &  & +\lambda_{2}\left(\text{Tr}[\tilde{\Phi}\Phi^{\dagger}]^{2}+\text{Tr}[\tilde{\Phi}^{\dagger}\Phi]^{2}\right)+\lambda_{3}\text{Tr}[\tilde{\Phi}\Phi^{\dagger}]\text{Tr}[\tilde{\Phi}^{\dagger}\Phi]+\lambda_{4}\text{Tr}[\Phi^{\dagger}\Phi]\left(\text{Tr}[\tilde{\Phi}\Phi^{\dagger}]+\text{Tr}[\tilde{\Phi}^{\dagger}\Phi]\right)\nonumber \\
 &  & +\rho_{1}\left(\text{Tr}[\Delta_{L}\Delta_{L}^{\dagger}]^{2}+\text{Tr}[\Delta_{R}\Delta_{R}^{\dagger}]^{2}\right)+\rho_{2}\left(\text{Tr}[\Delta_{L}\Delta_{L}]\text{Tr}[\Delta_{L}^{\dagger}\Delta_{L}^{\dagger}]+\text{Tr}[\Delta_{R}\Delta_{R}]\text{Tr}[\Delta_{R}^{\dagger}\Delta_{R}^{\dagger}]\right)\nonumber \\
 &  & +\rho_{3}\text{Tr}[\Delta_{L}\Delta_{L}^{\dagger}]\text{Tr}[\Delta_{R}\Delta_{R}^{\dagger}]+\rho_{4}\left(\text{Tr}[\Delta_{L}\Delta_{L}]\text{Tr}[\Delta_{R}^{\dagger}\Delta_{R}^{\dagger}]+\text{Tr}[\Delta_{L}^{\dagger}\Delta_{L}^{\dagger}]\text{Tr}[\Delta_{R}\Delta_{R}]\right) \\
 &  & +\alpha_{1}\text{Tr}[\Phi^{\dagger}\Phi]\left(\text{Tr}[\Delta_{L}\Delta_{L}^{\dagger}]+\text{Tr}[\Delta_{R}\Delta_{R}^{\dagger}])+\alpha_{3}(\text{Tr}[\Phi\Phi^{\dagger}\Delta_{L}\Delta_{L}^{\dagger}]+\text{Tr}[\Phi^{\dagger}\Phi\Delta_{R}\Delta_{R}^{\dagger}]\right)\nonumber \\
 &  & +\alpha_{2}\left(\text{Tr}[\Delta_{L}\Delta_{L}^{\dagger}]\text{Tr}[\tilde{\Phi}\Phi^{\dagger}]+\text{Tr}[\Delta_{R}\Delta_{R}^{\dagger}]\text{Tr}[\tilde{\Phi}^{\dagger}\Phi]+{\rm H.c.}\right) \nonumber \\
 &  & +\beta_{1}\left(\text{Tr}[\Phi\Delta_{R}\Phi^{\dagger}\Delta_{L}^{\dagger}]+\text{Tr}[\Phi^{\dagger}\Delta_{L}\Phi\Delta_{R}^{\dagger}]\right)+\beta_{2}\left(\text{Tr}[\tilde{\Phi}\Delta_{R}\Phi^{\dagger}\Delta_{L}^{\dagger}]+\text{Tr}[\tilde{\Phi}^{\dagger}\Delta_{L}\Phi\Delta_{R}^{\dagger}]\right)\nonumber \\
 &  & +\beta_{3}\left(\text{Tr}[\Phi\Delta_{R}\text{\ensuremath{\tilde{\Phi}^{\dagger}\Delta_{L}^{\dagger}}}]+\text{Tr}[\Phi^{\dagger}\Delta_{L}\text{\ensuremath{\tilde{\Phi}\Delta_{R}^{\dagger}}}]\right) \nonumber \,
\end{eqnarray}
where all couplings are assumed real. Here, $\tilde{\Phi}= \sigma_2 \Phi^*\sigma_2$, where $\sigma_2$ is the 2nd Pauli matrix. $\tilde{\Phi}$ transforms the same way as $\Phi$ does.
\par Assume that after the SSB, the vacuum state of the potential is stable and has the form of VEV structure eq. \eqref{eq:LRV-8}. We can then minimize the potential w.r.t the VEV parameters,
$$ \frac{\partial V}{\partial \kappa_1}=\frac{\partial V}{\partial \kappa_2}=\frac{\partial V}{\partial \theta_2}=\frac{\partial V}{\partial v_L}=\frac{\partial V}{\partial \theta_L}=\frac{\partial V}{\partial v_R}=0$$
This yields a set of 6 equations which can be solved to yield the famous VEV see-saw relation \cite{Dev:2018foq}.
\begin{equation}
    \beta_{1}\cos\left(\theta_{2}-\theta_{L}\right)\kappa_{2}\kappa_{1}+\beta_{2}\kappa_{1}^{2}\cos \theta_{L}+\beta_{3}\cos\left(2\theta_{2}-\theta_{L}\right)\kappa_{2}^{2}=\left(2\rho_{1}-\rho_{3}\right)v_{L}v_{R}
\end{equation}
Note if $\beta_{1,2,3}=0$ and since phenomenologically $v_R\neq0$, this implies $v_L=0$. 

\section{Vacuum Stability}\label{VS}

Quartic terms containing only the scalar bidoublet Higgs field constitutes the $\lambda$ sector and those containing only left and right-handed triplet Higgs fields constitutes the $\rho$ sector. It should be noted that mixing terms (i.e. involving $\alpha$'s and $\beta$'s) complicate the analysis for boundedness. We first look at bidoublet and triplets part of the potential separately to understand the procedure of minimization and useful parametrization to obtain BFB conditions. We then analyze the BFB condition for the potential in presence of non-zero quartic terms that couple bidoublet and triplet fields together in Sec \ref{dread}.

\subsection{Bidoublet $\Phi$ : $\lambda$ Sector}\label{subL}
As the potential should be bounded in all field directions, we first choose to find conditions for $\lambda$ sector containing the bidoublet $\Phi$. Considering only the quartic part, we require
\begin{eqnarray}\label{eq:potL}
 V^{\lambda}_{4} & \ = \ &  \lambda_{1}\text{Tr}[\Phi^{\dagger}\Phi]^{2}
  +\lambda_{2}\left(\text{Tr}[\tilde{\Phi}\Phi^{\dagger}]^{2}+\text{Tr}[\tilde{\Phi}^{\dagger}\Phi]^{2}\right)+\lambda_{3}\text{Tr}[\tilde{\Phi}\Phi^{\dagger}]\text{Tr}[\tilde{\Phi}^{\dagger}\Phi] \\
  & & +\lambda_{4}\text{Tr}[\Phi^{\dagger}\Phi]\left(\text{Tr}[\tilde{\Phi}\Phi^{\dagger}]+\text{Tr}[\tilde{\Phi}^{\dagger}\Phi]\right) >0\quad \forall \,\Phi\nonumber 
\end{eqnarray}
To obtain the conditions to be BFB, we parametrize $V^{\lambda}_{4}$ as follows:

\begin{eqnarray*}
\text{Tr}[\Phi^{\dagger}\Phi] & \ \equiv \ &   r^2 \\
\text{Tr}[\tilde{\Phi}\Phi^{\dagger}]/\text{Tr}[\Phi^{\dagger}\Phi] & \ \equiv \ & \xi e^{i \omega}\\
\text{Tr}[\tilde{\Phi}^{\dagger}\Phi]/\text{Tr}[\Phi^{\dagger}\Phi] & \ \equiv \ & \xi e^{-i \omega}
\end{eqnarray*}
where $r>0$, $\xi \in [0,1]$ and $\omega \in [0,2\pi]$. Quartic field terms present in the potential are normalized with the norm of the bidoublet $\Phi$ as discussed in sec \ref{GOS} .The complex product \text{Tr}[$\tilde{\Phi}\Phi^{\dagger}$]/\text{Tr}[$\Phi^{\dagger}\Phi$] between two unit spinors will be a complex number and hence has been parametrized accordingly. This approach to parametrization has been earlier used for obtaining boundedness criteria in two-Higgs-doublet Model \cite{ElKaffas:2006gdt,Ivanov:2006yq} and doublet-triplet-Higgs Model \cite{Arhrib:2011uy}. \newline
Substituting above in eq. \eqref{eq:potL},
\begin{equation}\label{eq:potLP}
 V^{\lambda}_{4}  =  r^4 \left( \lambda_{1}
  +2 \lambda_{2}\xi^2\cos{2\omega}
  +\lambda_{3}\xi^2 +2\lambda_{4}\,\xi\cos{\omega} \right)\equiv r^4 f(\lambda,\xi,\omega)
\end{equation}
We know from the extremum value theorem, the minimum of $V^{\lambda}_{4}$ must exist in/on the closed boundary defined by the disk. Furthermore, it should either exist inside the bounded region or on the boundary. We first minimize $V^{\lambda}_{4}$ inside the boundary w.r.t $\xi$ and $\omega$.

$$ f_\xi = \frac{\partial f}{\partial \xi} = 4 \lambda_{2}\xi\cos{2\omega} +2\lambda_{3}\xi
   +2\lambda_{4}\,\cos{\omega} =0 $$
$$ f_\omega = \frac{\partial f}{\partial \omega} = -4 \lambda_{2}\xi^2\sin{2\omega}
   -2\lambda_{4}\,\xi\sin{\omega} = -2\xi\sin{\omega}(4\lambda_{2}\xi\cos{\omega}+\lambda_{4})=0$$   
Here, we denote $\frac{\partial f}{\partial x}$ as $f_x$ and continue using this notation for conciseness. Solving the above two equations simultaneously, we get three critical points. Only the first two critical points are valid solutions of these pair of equations.\newline
$$f_\omega =0 \implies \xi=0, \, \sin{\omega}=0 \text{ or } \cos{\omega}=-\frac{\lambda_4}{4\lambda_2 \xi}$$\newline
\textbf{Case 1:} $\xi=0$
$$f_\xi = 2 \lambda_4 \cos{\omega}=0$$
$$\implies \cos{\omega}=0$$
Using this $\xi$ and $ \cos{\omega}$ in \eqref{eq:potLP}, we obtain the trivial condition for boundedness
\begin{equation}\label{eq:cl1}
    \lambda_1>0
\end{equation}
\newline
\textbf{Case 2:} $\sin{\omega}=0$ \newline
Notice, $\sin{\omega}=0 \implies \cos{\omega}=\pm 1$. From eq. \eqref{eq:potLP}, we notice this minimum value of $\cos{\omega}$ depends on the sign on $\lambda_4$. 
$$ \cos{\omega}= -\text{sgn}(\lambda_4)$$
Here, sgn(x) is the signum function. Thus, the relevant equation for minimum can be written as:
$$ f_\xi = 4 \lambda_{2}\xi +2\lambda_{3}\xi
   -2|\lambda_{4}| =0 $$
$$\implies \xi=\frac{|\lambda_4|}{2\lambda_2+\lambda_3}$$
Inserting these values in $f$ requiring $V^{\lambda}_{4}>0$, we get
$$\lambda_1+(2 \lambda_{2} +\lambda_{3})\left({\frac{|\lambda_4|}{2\lambda_2+\lambda_3}}\right)^{2}
 -2|\lambda_{4}|\, \frac{|\lambda_4|}{2\lambda_2+\lambda_3}>0$$
Thus, we get second condition as requirement:
\begin{equation}\label{eq:cl2}
   \lambda_1 -\frac{\lambda_4^2}{2\lambda_2+\lambda_3}  > 0 \quad \Longleftarrow \quad 2\lambda_2+\lambda_3 >   \left\lvert\lambda_4\right\rvert
\end{equation}
\newline
\textbf{Case 3:} $\cos{\omega}=-\frac{\lambda_4}{4\lambda_2 \xi}$
\newline
$$ 4 \lambda_{2}\xi\left(2\left(\frac{\lambda_4}{4\lambda_2 \xi} \right)^2 -1\right)
  +2\lambda_{3}\xi -2\lambda_{4}\left(\frac{\lambda_4}{4\lambda_2 \xi} \right)= -2\lambda_{3}\xi=0$$
The solution for above is $\xi=0$ but $\cos{\omega}$ is not defined for this value. Thus, this is not a valid solution.\\ \\
Now, we try to minimize $f$ on the boundary w.r.t to $\omega$ by setting $\xi=1$. 
$$ f_\omega = -4 \lambda_{2}\sin{2\omega}
   -2\lambda_{4}\,\sin{\omega} = -2\sin{\omega}(4\lambda_{2}\cos{\omega}+\lambda_{4})=0$$
\newline
\textbf{Case 4:} $\xi=1,\,\sin{\omega}=0 \implies \cos{\omega}=-\text{sgn}(\lambda_4),\quad \cos{2\omega}=1 $ \\
Using this we have the condition,
\begin{equation}\label{eq:cl3}
\lambda_{1} +\lambda_{3}
  +2( \lambda_{2}
 -|\lambda_{4}| )>0 
\end{equation}
\newline
\textbf{Case 5:} $\xi=1,\,\cos{\omega}=-\frac{\lambda_4}{4\lambda_2 }$
\newline
$$ \lambda_{1} + 2 \lambda_{2}\left(2\left(\frac{\lambda_4}{4\lambda_2} \right)^2 -1\right)
  +\lambda_{3} -2\lambda_{4}\left(\frac{\lambda_4}{4\lambda_2 \xi} \right)>0$$
The final condition can be written as:
\begin{equation}\label{eq:cl4}
 \lambda_{1} +\lambda_{3} -2 \lambda_{2}
   -\frac{\lambda_4^2}{4\lambda_2 }>0  \quad \Longleftarrow \quad  \left\lvert\frac{\lambda_4}{4\lambda_2}\right\rvert < 1 
\end{equation}

Thus, equations \eqref{eq:cl1}, \eqref{eq:cl2}, \eqref{eq:cl3} and \eqref{eq:cl4} collectively form the required  bounded from below (BFB) conditions for $\lambda$ sector. 
\par Now, we'll remark on the behaviour of these conditions to understand their characteristics in the plots. The condition with the minimum value dominates the boundedness of the potential. All conditions dominate in different regions of the parameter space and controls the boundedness of the potential. For instance, the condition from inside the boundary eq. \eqref{eq:cl2} dominates over other conditions if $2\lambda_2+\lambda_3 >   \left\lvert\lambda_4\right\rvert$ is satisfied. It can also be shown that eq. \eqref{eq:cl4} dominates when $\lambda_2>0$ otherwise eq. \eqref{eq:cl3} is valid.

\subsection{Triplets $\Delta_{L}$ and $\Delta_{R}$ : $\rho$ sector}\label{subR}

The quartic part of the potential with $\rho_i$'s is :

\begin{eqnarray}\label{eq:potR}
 V^{\rho}_{4} & \ = \ & \rho_{1}\left(\text{Tr}[\Delta_{L}\Delta_{L}^{\dagger}]^{2}+\text{Tr}[\Delta_{R}\Delta_{R}^{\dagger}]^{2}\right)+\rho_{2}\left(\text{Tr}[\Delta_{L}\Delta_{L}]\text{Tr}[\Delta_{L}^{\dagger}\Delta_{L}^{\dagger}]+\text{Tr}[\Delta_{R}\Delta_{R}]\text{Tr}[\Delta_{R}^{\dagger}\Delta_{R}^{\dagger}]\right)\nonumber \\
 &  & +\rho_{3}\text{Tr}[\Delta_{L}\Delta_{L}^{\dagger}]\text{Tr}[\Delta_{R}\Delta_{R}^{\dagger}]+\rho_{4}\left(\text{Tr}[\Delta_{L}\Delta_{L}]\text{Tr}[\Delta_{R}^{\dagger}\Delta_{R}^{\dagger}]+\text{Tr}[\Delta_{L}^{\dagger}\Delta_{L}^{\dagger}]\text{Tr}[\Delta_{R}\Delta_{R}]\right) \nonumber \\
\end{eqnarray}
To obtain the conditions for BFB, we parametrize $V^{\rho}_{4}$ similar to sec \ref{subL} :
\begin{eqnarray*}\label{eq:potRP}
\text{Tr}[\Delta_{L}\Delta_{L}^{\dagger}]+\text{Tr}[\Delta_{R}\Delta_{R}^{\dagger}]
& \ \equiv \ &   r^2 \\
\text{Tr}[\Delta_{L}\Delta_{L}^{\dagger}] & \ \equiv \ &   r^2 \sin^2 \gamma\\
\text{Tr}[\Delta_{R}\Delta_{R}^{\dagger}] & \ \equiv \ &   r^2 \cos^2 \gamma\\
\text{Tr}[\Delta_{L}\Delta_{L}]/\text{Tr}[\Delta_{L}\Delta_{L}^{\dagger}] & \ \equiv \ & \eta_1 e^{i \theta_1} \\
\text{Tr}[\Delta_{L}^{\dagger}\Delta_{L}^{\dagger}]/\text{Tr}[\Delta_{L}\Delta_{L}^{\dagger}] & \ \equiv \ & \eta_1 e^{-i \theta_1}\\
\text{Tr}[\Delta_{R}\Delta_{R}]/\text{Tr}[\Delta_{R}\Delta_{R}^{\dagger}] & \ \equiv \ & \eta_2 e^{i \theta_2} \\
\text{Tr}[\Delta_{R}^{\dagger}\Delta_{R}^{\dagger}]/\text{Tr}[\Delta_{R}\Delta_{R}^{\dagger}] & \ \equiv \ & \eta_2 e^{-i \theta_2}
\end{eqnarray*}
where $r>0$ , $\gamma \in [0,\frac{\pi}{2}]$, $\eta_1,\eta_2 \in [0,1]$ and $\theta_1,\theta_2 \in [0,2\pi]$.
Substituting above in eq. \eqref{eq:potR},
\begin{eqnarray}\label{eq:potRP}
 V^{\rho}_{4} & \ = \ & r^4(\rho_{1}\left(\cos^4 \gamma + \sin^4 \gamma\right)+\rho_{2}\left( \eta_1^2\sin^4 \gamma+ \eta_2^2 \cos^4 \gamma\right)\nonumber \\
     &  & +\rho_{3} \cos^2 \gamma  \sin^2 \gamma +2\rho_{4}\eta_1\eta_2 \cos(\theta_1-\theta_2)\cos^2 \gamma  \sin^2 \gamma) \equiv g(\rho,\gamma,\eta_{1,2},\theta_{1,2}) 
\end{eqnarray}
For minimum w.r.t to $\theta_1,\theta_2$ and taking in account sign of $\rho_4$, this can be rewritten as:
\begin{eqnarray}\label{eq:RP}
 V^{\rho}_{4} & \ = \ & \frac{r^4}{(1+\tan^2 \gamma)^2} \left(\tan^4 \gamma\left(\right.\rho_{1} + \rho_{2}\eta_1^2\right)+\tan^2\ \gamma \left(\rho_{3} -2|\rho_{4}|\eta_1\eta_2 \right) +\left.\rho_{1} + \rho_{2}\eta_2^2\right) \nonumber 
\end{eqnarray}
Requiring the above expression to be positive for all values of $\tan{\gamma}$ can be translated to $V^{\rho}_{4}$ being copositive for variable $\tan^2{\gamma}$. Thus, we have following requirements for $V^{\rho}_{4}$ to be bounded from below :
\begin{equation}\label{cr1}
\rho_{1} + \rho_{2}\eta_1^2>0 
\end{equation}
\begin{equation}\label{cr2}
\rho_{1} + \rho_{2}\eta_2^2>0    
\end{equation}
\begin{equation}\label{cr3}
\mathcal{G}(\rho,\eta_{1,2})\equiv\rho_{3} -2|\rho_{4}|\eta_1\eta_2 + 2\sqrt{(\rho_{1} + \rho_{2}\eta_1^2)(\rho_{1} + \rho_{2}\eta_2^2)} > 0   
\end{equation}
in regions  $\eta_1,\eta_2 \in [0,1]$.
\par Eq. \eqref{cr1} is equivalent to \eqref{cr2} as they are uncoupled in the constraint variable. Minimum value for the expression occurs at the endpoint as its monotonic in the quantity $\eta_i^2$, which ranges from $[0,1]$. Plugging the end points of the range of $\eta_i^2$, 
\begin{equation}\label{eq:RC1}
\rho_1 >0 
\end{equation}
\begin{equation}\label{eq:RC2}
\quad \rho_1  + \rho_2 >0
\end{equation}
We can first minimize $\mathcal{G}$ inside the boundary of square formed by $\eta_1$ and $\eta_2$. By minimizing the condition w.r.t to $\eta$'s, 
$$\mathcal{G}_{\eta_1}\equiv 2\eta_1\rho_2\frac{\sqrt{(\rho_{1} + \rho_{2}\eta_2^2)}}{\sqrt{(\rho_{1} + \rho_{2}\eta_1^2)}}-2\eta_2|\rho_4|=0 $$
$$\mathcal{G}_{\eta_2}\equiv 2\eta_2\rho_2\frac{\sqrt{(\rho_{1} + \rho_{2}\eta_1^2)}}{\sqrt{(\rho_{1} + \rho_{2}\eta_2^2)}}-2\eta_1|\rho_4|=0 $$
Solving the above two equations, we get
$$(\eta_1,\eta_2) = (0,0)$$  
Plugging it back in $\mathcal{G}$,
\begin{equation}\label{eq:RC3}
\rho_{3} + 2 \rho_{1}   > 0
\end{equation}
For minimizing $\mathcal{G}$ on the boundary, we set $\eta_1=\eta_2=1$. We obtain the condition
\begin{equation}\label{eq:RC4}
\rho_{3} - 2|\rho_{4}|+ 2 (\rho_{1} + \rho_{2})   > 0
\end{equation}
It can be proved that condition obtained by setting $\eta_1=0,\eta_2=1$ or vice-versa, always lies between the above two conditions and need not be checked for boundedness. Thus, conditions \eqref{eq:RC1}, \eqref{eq:RC2}, \eqref{eq:RC3} and \eqref{eq:RC4} collectively form the required conditions for BFB $\rho$ sector. 

\subsection{Dreaded Coupled Case: $\alpha_{1,3}\neq 0$}\label{dread}

This section outlines the procedure to find boundedness in presence of terms that couple the bidoublet and the triplet Higgs fields. For VEV see-saw relation to work naturally, we assume $\beta_i=0$  \cite{Maiezza:2016ybz}. This would imply $v_L=0$ and a non-zero $v_R$. Also $\alpha_2$ does not explicitly appears in the expressions for scalar mass spectrum. This gives us the freedom to set it to 0 for our analysis \cite{Dev:2016dja}. Thus, only $\alpha_1$ and $\alpha_3$ are assumed to be non-zero as they contribute to the scalar masses and have lower bounds on them from experimental constraints. The quartic part of the potential is given below:
\begin{eqnarray}
V_4 & \ = \  & \lambda_{1}\text{Tr}[\Phi^{\dagger}\Phi]^{2}\nonumber  +\lambda_{2}\left(\text{Tr}[\tilde{\Phi}\Phi^{\dagger}]^{2}+\text{Tr}[\tilde{\Phi}^{\dagger}\Phi]^{2}\right)+\lambda_{3}\text{Tr}[\tilde{\Phi}\Phi^{\dagger}]\text{Tr}[\tilde{\Phi}^{\dagger}\Phi]+\lambda_{4}\text{Tr}[\Phi^{\dagger}\Phi]\left(\text{Tr}[\tilde{\Phi}\Phi^{\dagger}]+\text{Tr}[\tilde{\Phi}^{\dagger}\Phi]\right)\nonumber \\
 &  & +\rho_{1}\left(\text{Tr}[\Delta_{L}\Delta_{L}^{\dagger}]^{2}+\text{Tr}[\Delta_{R}\Delta_{R}^{\dagger}]^{2}\right)+\rho_{2}\left(\text{Tr}[\Delta_{L}\Delta_{L}]\text{Tr}[\Delta_{L}^{\dagger}\Delta_{L}^{\dagger}]+\text{Tr}[\Delta_{R}\Delta_{R}]\text{Tr}[\Delta_{R}^{\dagger}\Delta_{R}^{\dagger}]\right)\nonumber \\
 &  & +\rho_{3}\text{Tr}[\Delta_{L}\Delta_{L}^{\dagger}]\text{Tr}[\Delta_{R}\Delta_{R}^{\dagger}]+\rho_{4}\left(\text{Tr}[\Delta_{L}\Delta_{L}]\text{Tr}[\Delta_{R}^{\dagger}\Delta_{R}^{\dagger}]+\text{Tr}[\Delta_{L}^{\dagger}\Delta_{L}^{\dagger}]\text{Tr}[\Delta_{R}\Delta_{R}]\right) \nonumber \\
 &  & +\alpha_{1}\text{Tr}[\Phi^{\dagger}\Phi]\left(\text{Tr}[\Delta_{L}\Delta_{L}^{\dagger}]+\text{Tr}[\Delta_{R}\Delta_{R}^{\dagger}]\right)+\alpha_3\left(\text{Tr}[\Phi\Phi^{\dagger}\Delta_{L}\Delta_{L}^{\dagger}]+ \text{Tr}[\Phi^{\dagger}\Phi\Delta_{R}\Delta_{R}^{\dagger}] \right)\label{eq:potA}
\end{eqnarray}
The parametrization in this case follows similarly as before. This has 3 different field directions and therefore can be parametrized on a sphere.
\begin{eqnarray*}
\text{Tr}[\Phi^{\dagger}\Phi]+\text{Tr}[\Delta_{L}\Delta_{L}^{\dagger}]+\text{Tr}[\Delta_{R}\Delta_{R}^{\dagger}]
& \ \equiv \ &   r^2 \\
\text{Tr}[\Phi^{\dagger}\Phi] & \ \equiv \ &   r^2 \cos^2 \theta \\
\text{Tr}[\Delta_{L}\Delta_{L}^{\dagger}] & \ \equiv \ &   r^2 \sin^2 \gamma  \sin^2 \theta\\
\text{Tr}[\Delta_{R}\Delta_{R}^{\dagger}] & \ \equiv \ &   r^2 \cos^2 \gamma  \sin^2 \theta\\
\text{Tr}[\tilde{\Phi}\Phi^{\dagger}]/\text{Tr}[\Phi^{\dagger}\Phi] & \ \equiv \ & \xi e^{i \omega} \\
\text{Tr}[\tilde{\Phi}^{\dagger}\Phi]/\text{Tr}[\Phi^{\dagger}\Phi] & \ \equiv \ & \xi e^{-i \omega} \\
\text{Tr}[\Delta_{L}\Delta_{L}]/\text{Tr}[\Delta_{L}\Delta_{L}^{\dagger}] & \ \equiv \ & \eta_1 e^{i \theta_1} \\
\text{Tr}[\Delta_{L}^{\dagger}\Delta_{L}^{\dagger}]/\text{Tr}[\Delta_{L}\Delta_{L}^{\dagger}] & \ \equiv \ & \eta_1 e^{-i \theta_1}\\
\text{Tr}[\Delta_{R}\Delta_{R}]/\text{Tr}[\Delta_{R}\Delta_{R}^{\dagger}] & \ \equiv \ & \eta_2 e^{i \theta_2} \\
\text{Tr}[\Delta_{R}^{\dagger}\Delta_{R}^{\dagger}]/\text{Tr}[\Delta_{R}\Delta_{R}^{\dagger}] & \ \equiv \ & \eta_2 e^{-i \theta_2}\\
\text{Tr}[\Phi\Phi^{\dagger}\Delta_{L}\Delta_{L}^{\dagger}]/\text{Tr}[\Phi^{\dagger}\Phi]\text{Tr}[\Delta_{L}\Delta_{L}^{\dagger}] & \ \equiv \ & \zeta_1 \\
\text{Tr}[\Phi^{\dagger}\Phi\Delta_{R}\Delta_{R}^{\dagger}]/\text{Tr}[\Phi^{\dagger}\Phi]\text{Tr}[\Delta_{R}\Delta_{R}^{\dagger}] & \ \equiv \ & \zeta_2 
\end{eqnarray*}
with $r>0$, $|\xi|\leq1$, $\theta \in [0,\frac{\pi}{2}]$, $\gamma \in [0,\frac{\pi}{2}]$, $\eta_1,\eta_2 \in [0,1]$ , $\theta_1,\theta_2 \in [0,2\pi]$ \\
\begin{figure}

\includegraphics[width=0.6\textwidth]{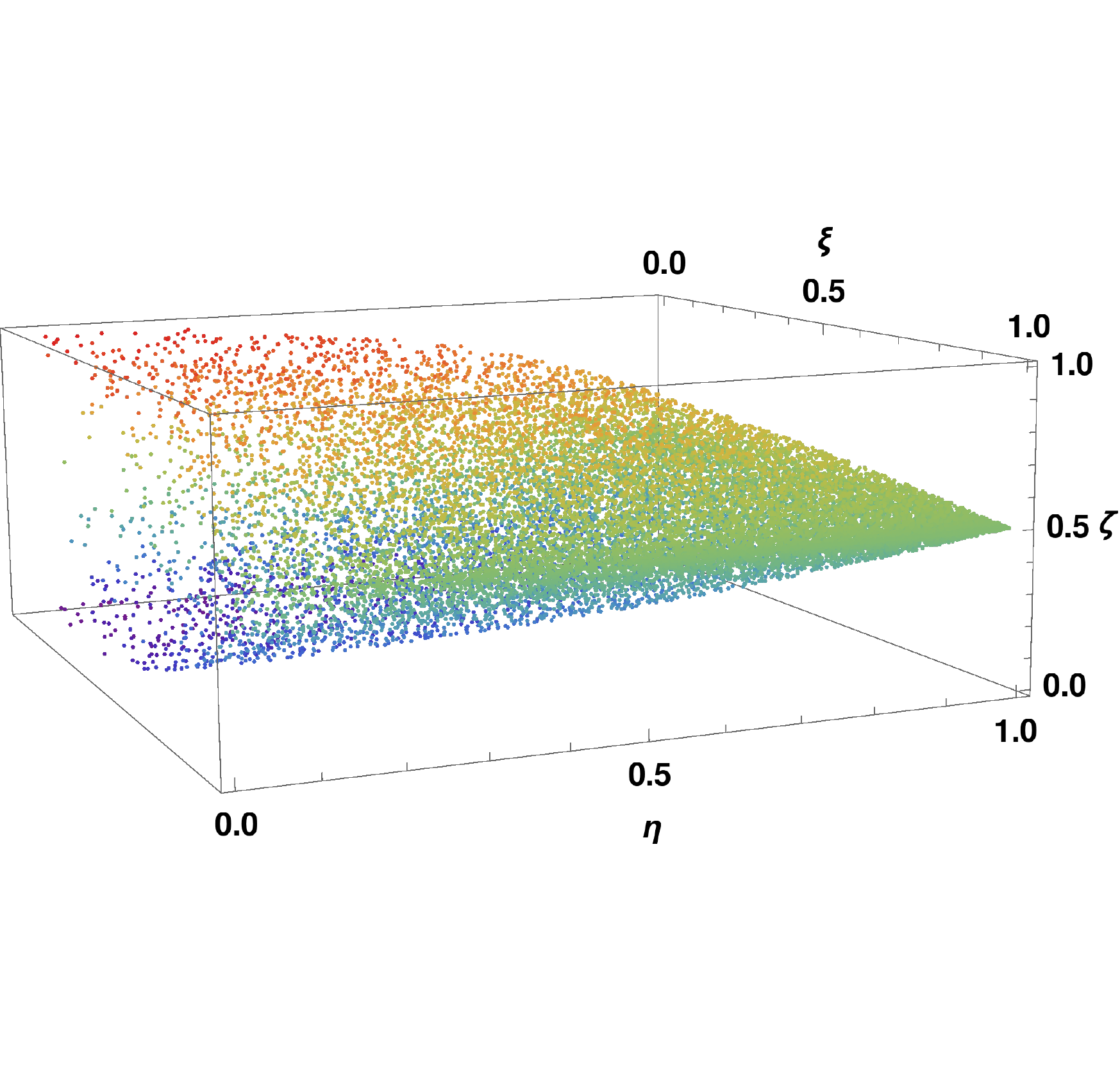}\ \includegraphics[width=0.38\textwidth]{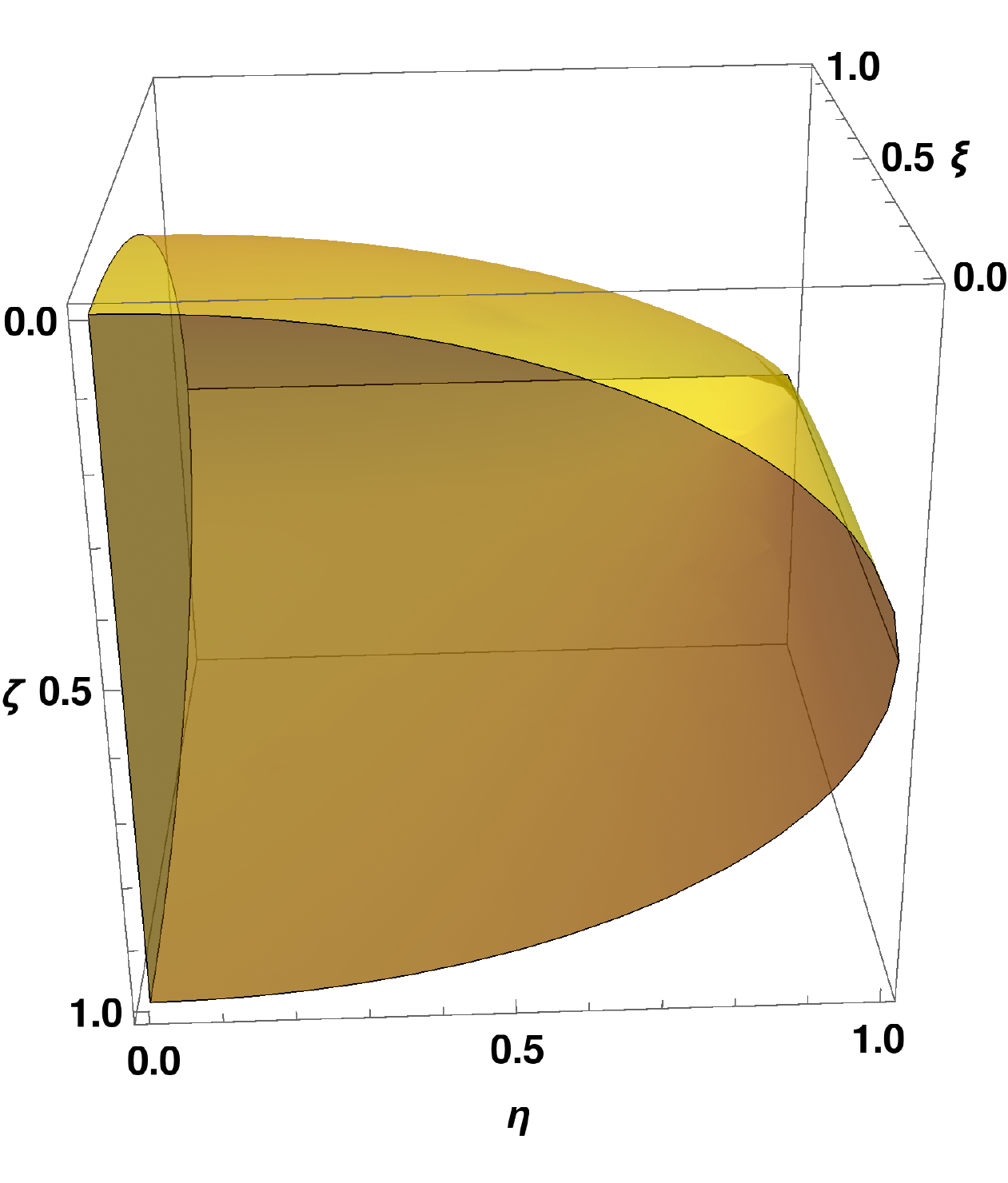}

\caption{\label{fig:relatedZeta}Dependence of gauge orbit variable $\zeta_i$ on $\xi$ and $\eta_i$. (Left) Scatter plot of $\zeta$ with respect to $\xi$ and $\eta$. (Right) Plot of $\zeta$ as a function of $\xi$ and $\eta$ given in eq. \eqref{eq:zeta}.}
\end{figure}
\par Naively, it might be expected that $\zeta_1,\zeta_2 \in [0,1]$ \cite{Bonilla:2015eha}. However, as can be seen from the scatter plot in Fig \ref{fig:relatedZeta}, $\zeta_i$ depends on $\xi_i$ and $\eta_i$. In fact, it can be shown that value of $\zeta_i$ is bounded from above and below given by,
\begin{equation}
    \frac{1}{2}\left(1-\sqrt{1-\xi^2}\sqrt{1-\eta_i^2}\right)\leq \zeta_i \leq \frac{1}{2}\left(1+\sqrt{1-\xi^2}\sqrt{1-\eta_i^2}\right)
    \label{eq:zeta}
\end{equation}
where $i\in \{1,2\}$, $|\xi|\leq1$ and  $\eta_i \in [0,1]$. As can be seen in Fig \ref{fig:relatedZeta}, the dependence of $\zeta_i$ on $\xi$ and $\eta_i$ depicted in the scatter plot is captured exactly in eq. \eqref{eq:zeta}. 
\\ \\
Substituting the above defined gauge orbit variables in eq. \eqref{eq:potA},
\begin{eqnarray}\label{eq:potAP}
 V_{4} & \ = \  & r^4 \cos^4 \theta \left( \lambda_{1}\nonumber 
  +2 \lambda_{2}\xi^2\cos{2\omega}
  +\lambda_{3}\xi^2 +2\lambda_{4}\,\xi\cos{\omega} \right)  \nonumber \\ 
 &  & +  r^4 \sin^4 \theta \left(\rho_{1}\left(\cos^4 \gamma + \sin^4 \gamma\right)+\rho_{2}\left( \eta_1^2\sin^4 \gamma+ \eta_2^2 \cos^4 \gamma\right)\right.\nonumber \\
 &  & +\left.\rho_{3} \cos^2 \gamma  \sin^2 \gamma +2\rho_{4}\eta_1\eta_2 \cos(\theta_1-\theta_2)\cos^2 \gamma  \sin^2 \gamma\right)  \nonumber \\
 &  &+ \left(\alpha_1 +  \alpha_3(\zeta_1\cos^2\gamma + \zeta_2\sin^2\gamma)\right) r^4 \cos^2 \theta \sin^2 \theta  \nonumber \\ \nonumber \\
 & \equiv &  r^4 \left( \cos^4 \theta f(\lambda,\xi,\omega) +  \sin^4 \theta g(\rho,\gamma,\eta_{1,2},\theta_{1,2}) +  h(\alpha,\gamma,\zeta_{1,2}) \cos^2 \theta \sin^2 \theta \right) 
\end{eqnarray}
From copositivity criteria, it implies :
$$ f(\lambda,\xi,\omega) > 0 $$
$$ g(\rho,\gamma,\eta_{1,2},\theta_{1,2}) > 0$$
$$ h(\alpha,\gamma,\zeta_{1,2}) + 2\sqrt{f(\lambda,\xi,\omega)\:  g(\rho,\gamma,\eta_{1,2},\theta_{1,2})}>0$$
These conditions should hold for all values of $(\xi,\omega,\gamma,\eta_{1,2},\theta_{1,2},\zeta_{1,2})$. First two conditions are \eqref{eq:potLP} and \eqref{eq:potRP}, evaluated in previous sections. For 2nd and 3rd condition, minimum of $\theta_{1,2}$ can again be absorbed in the sign of $\lambda_4$.  
\begin{equation}
\alpha_1 +  \alpha_3(\zeta_1\cos^2\gamma + \zeta_2\sin^2\gamma) + \sqrt{f(\lambda,\xi,\omega)\:g(\rho,\gamma,\eta_{1,2})} > 0 
\end{equation}
where $f$ and $g$ can be written as :
\begin{equation*}
  f \equiv  \lambda_{1}\nonumber 
  +2 \lambda_{2}\xi^2\cos{2\omega}
  +\lambda_{3}\xi^2 +2\lambda_{4}\,\xi\cos{\omega}
\end{equation*}
\begin{equation}\label{eq:g}
g\equiv  \frac{1}{(1+\tan^2 \gamma)^2} \left(\tan^4 \gamma\left(\right.\rho_{1} + \rho_{2}\eta_2^2\right)+\tan^2\ \gamma \left(\rho_{3} -2|\rho_{4}|\eta_1\eta_2 \right) +\left.\rho_{1} + \rho_{2}\eta_1^2\right)
\end{equation}
We now turn to symmetries to simplify further and reduce minimizing variables. We first try to minimize condition $g(\rho,\gamma,\eta_{1,2})$ again but using symmetry arguments as an example. Note that $g$ is symmetric under the following operation:
$$ \cos\gamma \leftrightarrow \sin\gamma, \qquad \eta_1 \leftrightarrow \eta_2$$
Thus, the minimum occurs at $\cos\gamma = \sin\gamma$ and $\eta_1= \eta_2$, which yields :
$$g\equiv \frac{\rho_3+2\rho_1+2(\rho_2-|\rho_4|)\eta_1^2}{4}$$
Plugging in the endpoints for $\eta_1$, we obtain two conditions :
$$g=\left\{ \frac{\rho_3+2\rho_1}{4}, \frac{\rho_3+2\rho_1+2(\rho_2-|\rho_4|)}{4} \right\}$$
Previously, we minimized the condition for $\cos\gamma = \sin\gamma$ i.e. $\tan\gamma=1$. Now, we will minimize $g$ for the endpoint, $\tan\gamma=0$. Using the symmetry operations used above, it can be shown that minimizing the condition for other endpoint $\tan\gamma=\infty$ is equivalent to case for $\tan\gamma=0$. For this case, condition takes the form:
$$g\equiv \rho_1+2\rho_2\eta_1^2$$
Again plugging in the endpoints for $\eta_1$, we obtain two conditions. We obtain a total of 4 conditions for minimizing $g$, which exactly matches the conditions derived in sec. \ref{subR}.
\begin{equation}\label{eq:minG}
    g : \left\{ \rho_{1},\, \rho_{1} + \rho_{2},\, \frac{\rho_{3}+2\rho_{1}}{4},\, \frac{\rho_{3} - 2|\rho_{4}|+ 2 (\rho_{1} + \rho_{2})}{4}\right\}
\end{equation}
Note that the 3rd condition is symmetric under the following operation:
$$ \zeta_1 \leftrightarrow \zeta_2, \qquad \cos\gamma \leftrightarrow \sin\gamma, \qquad \eta_1 \leftrightarrow \eta_2$$
Thus, the function takes its minimum value inside the gauge orbit space when:
$$ \zeta_1 = \zeta_2, \qquad \cos\gamma = \sin\gamma, \qquad \eta_1= \eta_2$$
Using the above symmetry arguments, the form of the 3rd condition is :
\begin{equation}\label{eq:case1}
   \alpha_1 +  \alpha_3\zeta_1 + \sqrt{f(\lambda,\xi,\omega)\:(\rho_3+2\rho_1+2(\rho_2-|\rho_4|)\eta_1^2)} >0 
\end{equation}
For this case, the condition is monotonic in $\zeta_{1}$ \& $\eta_1^2$ and are trivially minimized at endpoints of their range. This implies for $\alpha_3 <0$, the most constraining condition corresponds to $\zeta_1 = \zeta_1^{max}$ and $\zeta_1 = \zeta_i^{min}$ for $\alpha_3 > 0$.
$$\zeta_i^{max}= \frac{1}{2}\left(1+\sqrt{1-\xi^2}\sqrt{1-\eta_i^2}\right),\quad \zeta_i^{min}= \frac{1}{2}\left(1-\sqrt{1-\xi^2}\sqrt{1-\eta_i^2}\right)$$
The minimum of $f$ has been evaluated in a previous section. This also yields corresponding value of $\xi$ and $\eta_1$ that determines the value of $\zeta_i^{max}$ and $\zeta_i^{min}$. This yields a set of 10 different conditions. \par Consider an example for above discussion. Let us assume  $f(\lambda,\xi,\omega)$ minimizes for $\xi=\frac{|\lambda_4|}{2\lambda_2+\lambda_3}$ and $\eta_1=0$, then $\zeta_1$ is given by :
$$\implies \zeta_1= \frac{1}{2}\left(1\pm\sqrt{1-\left(\frac{|\lambda_4|}{2\lambda_2+\lambda_3}\right)^2}\right) $$
then the required inequality to be checked for vacuum stability becomes :
$$ \alpha_1 + \frac{\alpha_3}{2}\left( 1\pm\sqrt{1-\frac{\lambda_4^2}{(2\lambda_2+\lambda_3)^2}}\right)+ \sqrt{\left(\lambda_1 -\frac{\lambda_4^2}{2\lambda_2+\lambda_3}\right)( \rho_{3}+ 2 \rho_{1} )}>0 $$
We also need to minimize the 3rd condition for the edge surface of $\tan\gamma$. For the case $\tan\gamma=0$, 3rd condition takes the form:
$$ \alpha_1 +  \alpha_3\zeta_1 + 2\sqrt{f(\lambda,\xi,\omega)\:(\rho_1+\rho_2\eta_1^2)} >0  $$
The above condition can be minimized similarly as in case of $\tan\gamma=1$, yielding a total of another 10 conditions.  
\par Thus, minimizing the 3rd condition yields a set of 20 inequalities to be checked. We have finally derived all conditions required for the vacuum stability of the LRSM. The complete set of these necessary and sufficient conditions are collected below :

\begin{empheq}[box=\widefbox]{gather}
\textbf{Analytic Conditions for Vacuum Stability in LRSM} \nonumber \\ \nonumber \\
{\rm f>0:\ }  \begin{cases}
 \lambda_{1}  \\
 \left(\lambda_1 -\frac{\lambda_4^2}{2\lambda_2+\lambda_3}\right) \qquad \quad \: \impliedby
 2\lambda_2+\lambda_3>|\lambda_4| \\
\left(\lambda_{1} +\lambda_{3}
+2( \lambda_{2}
-|\lambda_{4}|)\right)\\
\left(\lambda_{1} +\lambda_{3} -2 \lambda_{2}
-\frac{\lambda_4^2}{4\lambda_2 }\right) \impliedby |4\lambda_2|>|\lambda_4| \nonumber
\end{cases} \nonumber \\ \nonumber \\ \label{eq:VSCond}
  g > 0: \left\{ \rho_{1},\, \rho_{1} + \rho_{2},\, \frac{\rho_{3}+2\rho_{1}}{4},\, \frac{\rho_{3} - 2|\rho_{4}|+ 2 (\rho_{1} + \rho_{2})}{4}\right\} \\  \nonumber \\
\alpha_1 +  2\sqrt{\lambda_1\rho_{1}}>0 \nonumber \\
\alpha_1 + \alpha_3 +  2\sqrt{\lambda_1\rho_{1}}>0\nonumber \\
\alpha_1 +\frac{\alpha_3}{2}+ 2\sqrt{\lambda_1( \rho_{1} + \rho_{2})}>0\nonumber \\
\alpha_1 + \sqrt{\lambda_1( \rho_{3}+ 2 \rho_{1} )}>0\nonumber \\
\alpha_1 + \alpha_3+ \sqrt{\lambda_1( \rho_{3}+ 2 \rho_{1} )}>0\nonumber \\
\alpha_1 + \frac{\alpha_3}{2} + \sqrt{\lambda_1( \rho_{3} - 2|\rho_{4}|+ 2 (\rho_{1} + \rho_{2}))}>0\nonumber \\
\alpha_1 + \frac{\alpha_3}{2}\left( 1\pm\sqrt{1-\frac{\lambda_4^2}{(2\lambda_2+\lambda_3)^2}}\right) +  2\sqrt{\left(\lambda_1 -\frac{\lambda_4^2}{2\lambda_2+\lambda_3}\right) \rho_{1}}>0\nonumber \\
\alpha_1 + \frac{\alpha_3}{2}+ 2\sqrt{\left(\lambda_1 -\frac{\lambda_4^2}{2\lambda_2+\lambda_3}\right)( \rho_{1} + \rho_{2})}>0\nonumber \\
\alpha_1 + \frac{\alpha_3}{2}\left( 1\pm\sqrt{1-\frac{\lambda_4^2}{(2\lambda_2+\lambda_3)^2}}\right)+ \sqrt{\left(\lambda_1 -\frac{\lambda_4^2}{2\lambda_2+\lambda_3}\right)( \rho_{3}+ 2 \rho_{1} )}>0\nonumber \\
\alpha_1 + \frac{\alpha_3}{2}+ \sqrt{\left(\lambda_1 -\frac{\lambda_4^2}{2\lambda_2+\lambda_3}\right)( \rho_{3} - 2|\rho_{4}|+ 2 (\rho_{1} + \rho_{2}))}>0\nonumber \\
\alpha_1 +\frac{\alpha_3}{2} +  2\sqrt{\left( \lambda_{1} +\lambda_{3}
  -2 \lambda_{2}
  -\frac{\lambda_{4}^2}{4\lambda_2}\right)\: \textbf{Min}(g)}>0 \nonumber \\
\alpha_1 +\frac{\alpha_3}{2} +  2\sqrt{\left(\lambda_{1} +\lambda_{3}
  +2( \lambda_{2}
 -|\lambda_{4}| )\right)\: \textbf{Min}(g)}>0 \nonumber 
\end{empheq}
For using these conditions, we first ensure $f$ and $g$ should be strictly positive at all minima. For some conditions in $f$, we have the following structure $p\impliedby q$. This implies condition $p$ only needs to be checked if and only if condition $q$ is true. We then check rest of the conditions based on minimum value of $f$ and $g$.

\section{Symmetry Breaking and Desirable Vacuum}\label{SB}
A BFB potential does not necessarily leads to correct symmetry breaking yielding the correct ground state of the Higgs potential. Recently, some useful conditions (though not necessary) for a good vacuum in the left-right model were derived for a limited parameter space in \cite{Dev:2018foq}. Gauge-independent criteria to obtain a good vacuum was also proposed.
$$ \langle \Phi \rangle \neq 0 $$
 $$ \operatorname{det}\langle \Delta_R \rangle = \operatorname{det}\langle \Delta_L \rangle =  0  $$
$$ \langle \Delta_R \rangle \neq \langle \Delta_L \rangle$$
The first condition leads to non-zero expectation for Higgs VEV in the Standard Model. The second condition is required for $U(1)_{em}$ not to be broken. The third condition is required for broken parity at low energies. Although reference \cite{Dev:2018foq} specifies 4 conditions for a good vacuum but effectively only 3 conditions are required. As their condition $\langle \Delta_R \rangle \neq 0 \text{ or } \langle \Delta_L \rangle \neq 0$ for good vacuum is contained in $\langle \Delta_R \rangle \neq \langle \Delta_L \rangle$.   
\par In this section, we derive some useful conditions for scalar potential to exhibit correct spontaneous symmetry breaking (SSB) and specify the gauge-independent criteria for correct vacuum in more general form. Using the VEV structure of the scalar fields eq. \eqref{eq:LRV-8} in the general scalar potential eq. \eqref{eq:LRV-4}, 
\begin{eqnarray}\label{eq:Vssb}
V & \ = \ & -\frac{\left(\kappa _1^2+\kappa _2^2\right)}{2} \mu _1^2 -2 \kappa _1 \kappa _2 \mu _2^2 \cos (\theta_2)- \mu _3^2 \left(v_L^2+v_R^2\right) +\frac{\left(\kappa _1^2+\kappa _2^2\right)^2}{4} \lambda _1\\ \nonumber
&  &+ 2 \kappa _1^2 \kappa _2^2 \lambda_2 \cos (2 \theta_2 )+ \kappa _1 \kappa _2 \left(\kappa _1^2+\kappa _2^2\right) \lambda _4 \cos (\theta_2           )  + \kappa _1^2 \kappa _2^2 \lambda _3\\ \nonumber
&  &+ \rho _1 \left(v_L^4+v_R^4\right) + \rho _3 v_L^2 v_R^2 \\ \nonumber
&  & + \alpha _1 \frac{\left(\kappa _1^2+\kappa _2^2\right)}{2} \left(v_L^2+v_R^2\right) +\alpha _3 \frac{\kappa _2^2}{2} \left(v_L^2+v_R^2\right) 
\end{eqnarray}
For boundedness, the quartic part of the potential can be written as:
\begin{eqnarray}\label{eq:potSSB}
V_4 & \equiv &  r^4 \left(  f_{SSB}(\lambda,\xi,\omega)\cos^4 \theta+   g_{SSB}(\rho,\gamma,\theta_{1,2})\sin^4 \theta +  h_{SSB}(\alpha,\gamma,\zeta_{1,2}) \cos^2 \theta \sin^2 \theta \right)
\end{eqnarray}
where parametrizing variables are defined in accordance with section \ref{dread}. To obtain necessary and sufficient conditions for correct symmetry breaking, the minimum from the potential $V_{SSB}$ should be deeper than the one obtained from the general potential. Using eq. (\ref{eq:min}), the required condition can be written as :
\begin{equation}\label{eq:Weak}
-\frac{g\mu_1^4-h\mu_1^2\mu_2^2+f\mu_2^4}{4fg-h^2} >   -\frac{g_{SSB}\mu_1^4-h_{SSB}\mu_1^2\mu_2^2+f_{SSB}\mu_2^4}{4f_{SSB}\:g_{SSB}-h_{SSB}^2}
\end{equation}
The above relation needs to be minimized for the entire gauge orbit parameter space. Due to the non-linearity of the orbit variables, this is not analytically tractable. 
\par The important observation in this work is that the conditions sufficient for a general potential to lead to a good vacuum after SSB can be obtained by requiring VEV aligned scalar potential to dominate the general scalar potential i.e. $V\geq V_{SSB}$. This is a stronger condition than eq. (\ref{eq:Weak}) and using eq. \eqref{eq:potAP}, \eqref{eq:potSSB}  can be written as :
\begin{equation}
    (f-f_{SSB})\cos^4 \theta+   (g-g_{SSB})\sin^4 \theta +  (h-h_{SSB}) \cos^2 \theta \sin^2 \theta \geq 0
\end{equation}
Thus, for VEV structure in eq. \eqref{eq:LRV-8} to be the global minima of the theory, following conditions are required :
\begin{eqnarray}\label{fgSSB}
f\geq f_{SSB},\quad g\geq g_{SSB}
\end{eqnarray}
\begin{eqnarray}\label{hSSB}
 h-h_{SSB} + 2\sqrt{(f-f_{SSB})\:(g-g_{SSB})}\geq 0
\end{eqnarray}
It is also required that $V_{SSB}$ exhibits stable vaccum, which implies :
\begin{eqnarray}
 f_{SSB}>0,\quad  g_{SSB}>0,\quad
 h_{SSB} + 2\sqrt{f_{SSB}\:g_{SSB}}> 0
\end{eqnarray}
We begin by noticing that in eq. (\ref{eq:Vssb}), $f_{SSB}$ takes the same form as $f(\lambda,\xi,\omega)$ for the general potential. VEV condition $\langle\Phi\rangle\neq 0$ translates to $r\cos{\theta}\neq0$. It is satisfied as long as $\lambda$ sector is bounded from below. This implies all the conditions found for $\lambda$ sector are required for existence of a good vacuum. It also implies $f=f_{SSB}$ trivially satisfies condition for correct symmetry breaking.\\
On the other hand, $g_{SSB}$ has $\eta_{1,2}=0$. 
$$ \text{Tr}[\langle\Delta_{L}\rangle\langle\Delta_{L}\rangle]=0 \quad \implies \eta_1=0$$
$$ \text{Tr}[\langle\Delta_{R}\rangle\langle\Delta_{R}\rangle]=0 \quad \implies \eta_2=0$$
Therefore, coefficients of $\rho_2$ and $\rho_4$ vanish leading to following form of $g$ (See eq. \eqref{eq:g}):
$$ g_{SSB}\equiv  \frac{1}{(1+\tan^2 \gamma)^2} \left(\rho_{1} \tan^4 \gamma +\rho_{3} \tan^2\ \gamma  +\rho_{1}\right) $$
The minimum for this expression occurs at $\tan^2\gamma=0 \text{ or }1$. We require $\langle\Delta_{L}\rangle<\langle\Delta_{R}\rangle$ which can be easily shown equivalent to :
$$\text{Tr}[\langle\Delta_{L}\rangle\langle\Delta_{L}^{\dagger}\rangle]<\text{Tr}[\langle\Delta_{R}\rangle\langle\Delta_{R}^{\dagger}\rangle] $$
So according to the chosen parametrization, the preferred minima is $\tan^2 \gamma=0$. We know from sec. \ref{GOS}, condition with less positive value dominates the minima. Thus, this condition should dominate over the other minima i.e $\tan^2 \gamma=1$ in $g_{SSB}$. Thus, we require
$$ \frac{\rho_{3}+2\rho_{1}}{4}\geq\rho_1 \implies \rho_{3}-2\rho_{1} \geq 0$$
After requiring the internal structure of the VEV alignment, we want eq. \eqref{fgSSB} to hold i.e. $g \geq g_{SSB}$ should hold. The minimum of $g_{SSB}$ occurs for $\rho_1>0$. This condition should dominate other possible minima of the general potential. Using minimum conditions from eq. \eqref{eq:minG},
$$ \rho_1 +\rho_2 \geq \rho_1 \implies \rho_2\geq0$$
$$ \frac{\rho_{3}+2\rho_{1}}{4} \geq \rho_1 \implies \rho_{3}-2\rho_{1} \geq 0$$
$$ \frac{\rho_{3} - 2|\rho_{4}|+ 2 (\rho_{1} + \rho_{2})}{4} \geq \rho_1 \implies |\rho_{4}| \leq \frac{\rho_{3} - 2 \rho_{1} }{2}+ \rho_{2}  $$
Since $f=f_{SSB}$, eq. \eqref{hSSB} implies $h\geq h_{SSB}$. 
$$\alpha_1 +  \alpha_3(\zeta_1\cos^2\gamma + \zeta_2\sin^2\gamma) \geq \alpha_1 +  \alpha_3(\zeta_1^{SSB}\cos^2\gamma + \zeta_2^{SSB}\sin^2\gamma)$$ 
Note that since $\eta_i=0$ for $V_{SSB}$, $\zeta_i\neq\zeta_i^{SSB}$. The condition above is monotonic in $\zeta$ 's and the endpoints of their range can be substituted depending on the sign of $\alpha_3$.
$$\frac{\alpha_3}{2}\left(1-\textbf{Sgn}(\alpha_3)\sqrt{1-\xi^2}\sqrt{1-\eta_i^2} \right) \geq \frac{\alpha_3}{2}\left(1-\textbf{Sgn}(\alpha_3)\sqrt{1-\xi^2} \right)$$
As can be seen directly, the above condition holds true for all $\xi$, $\eta$ and $\alpha_3$. Similarly, for vacuum stability condition to hold true, we require :
\begin{equation}
   \alpha_1+ \frac{\alpha_3}{2}\zeta + 2\sqrt{f(\lambda,\xi,\omega)\:\rho_1} >0
\end{equation}
where, $\zeta= \left(1-\textbf{Sgn}(\alpha_3)\sqrt{1-\xi_*^2}\right)$ and $\xi_*$ equals the value of $\xi$ that minimizes $f(\lambda,\xi,\omega)$. The minimization of $f$ has been covered in sec. \ref{subL}.
\par For non-zero field values to be the global minimum (refer sec. \ref{GOS}), we also require eq. \eqref{muc1},\eqref{muc2} and \eqref{muc3} to hold. For non-zero $\langle\Phi\rangle$ and $\langle\Delta_{R}\rangle$, we require :
\begin{equation}
    2\:\textbf{Min}[f_{SSB}]\mu_3^2 - \textbf{Min}[h_{SSB}]\bar{\mu}_1^2 >0  \nonumber
\end{equation}
\begin{equation}
2\: \textbf{Min}[g_{SSB}]\bar{\mu}_1^2 - \textbf{Min}[h_{SSB}]\mu_3^2 >0 
\end{equation}
where,
$$\bar{\mu}_1^2 = \mu_1^2+2 \sigma  \mu_2^2, \quad \quad\sigma=\xi \cos\omega$$
and 
$$2\sqrt{\textbf{Min}[f_{SSB}]\:\textbf{Min}[g_{SSB}]} -\mathbf{||}\textbf{Min}[h_{SSB}]\mathbf{||}>0$$
Here, expression for $\bar{\mu}_1^2$ has been obtained by using parametrization from sec. (\ref{subL}) to relevant mass-squared terms in the scalar potential. Thus, the complete set of conditions sufficient to obtain a correct vacuum after SSB in left-right symmetric model are stated below:

\begin{empheq}[box=\widefbox]{gather}
\textbf{Analytic Conditions for Symmetry Breaking to Correct Vacuum } \nonumber \\  \nonumber \\ 
{\rm f_{SSB}>0:\ }  \begin{cases}
 \lambda_{1}>0, \quad \xi=\sigma=0\,,  \\
 \left(\lambda_1 -\frac{\lambda_4^2}{2\lambda_2+\lambda_3}\right)>0\impliedby 0<\xi=\frac{|\lambda_4|}{2\lambda_2+\lambda_3}
 <1, \quad 
 \sigma=- \frac{\lambda_4}{2\lambda_2+\lambda_3},\\
\left(\lambda_{1} +\lambda_{3}
+2( \lambda_{2}
-|\lambda_{4}|)\right)>0, \quad \xi=1,\:\sigma=-\textbf{Sgn}(\lambda_4)\,, \\
\left(\lambda_{1} +\lambda_{3} -2 \lambda_{2}
-\frac{\lambda_4^2}{4\lambda_2 }\right) >0\impliedby |4\lambda_2|>|\lambda_4|, \quad\xi=1,\: \sigma=-\frac{\lambda_4}{4\lambda_2 }, \nonumber
\end{cases} \nonumber \\ \nonumber \\
 \rho_1>0, \quad \rho_2>0, \quad \rho_{3}>2\rho_{1},\quad |\rho_{4}|< \frac{\rho_{3} - 2 \rho_{1} }{2}+ \rho_{2} \nonumber \\  \nonumber \\ \label{eq:SSBCond}
\alpha_1 +  2\sqrt{\lambda_1\rho_{1}}>0 \nonumber \\
\alpha_1 + \alpha_3 +  2\sqrt{\lambda_1\rho_{1}}>0 \nonumber \\
\alpha_1 + \frac{\alpha_3}{2}\left( 1\pm\sqrt{1-\frac{\lambda_4^2}{(2\lambda_2+\lambda_3)^2}}\right) +  2\sqrt{\left(\lambda_1 -\frac{\lambda_4^2}{2\lambda_2+\lambda_3}\right) \rho_{1}}>0\nonumber \\
\alpha_1 +\frac{\alpha_3}{2} +  2\sqrt{\left( \lambda_{1} +\lambda_{3}
  -2 \lambda_{2}
  -\frac{\lambda_{4}^2}{4\lambda_2}\right)\: \rho_1}>0 \nonumber \\
\alpha_1 +\frac{\alpha_3}{2} +  2\sqrt{\left(\lambda_{1} +\lambda_{3}
  +2( \lambda_{2}
 -|\lambda_{4}| )\right)\: \rho_1}>0 \nonumber \\ \nonumber\\  \bar{\mu}_1^2 = \mu_1^2+2 \sigma  \mu_2^2 \nonumber \\ \nonumber\\
 2\sqrt{\textbf{Min}[f_{SSB}]\:\rho_1} -\left|\left|\alpha_1 + \frac{\alpha_3}{2}\left( 1-\textbf{Sgn}(\alpha_3)\sqrt{1-\xi^2}\right) \right|\right|>0 \nonumber \\ \nonumber \\
 2\:\textbf{Min}[f_{SSB}]\mu_3^2 - \left[\alpha_1 + \frac{\alpha_3}{2}\left( 1-\textbf{Sgn}(\alpha_3)\sqrt{1-\xi^2}\right) \right]\bar{\mu}_1^2 >0  \nonumber \\
2 \rho_1\bar{\mu}_1^2 - \left[\alpha_1 + \frac{\alpha_3}{2}\left( 1-\textbf{Sgn}(\alpha_3)\sqrt{1-\xi^2}\right) \right]\mu_3^2 >0 \nonumber 
\end{empheq}
For using these conditions, we first ensure $f_{SSB}$  should be strictly positive at all minima. For some conditions in $f$, we have the following structure $p\impliedby q,\:\xi = \texttt{value1},\: \sigma = \texttt{value2}$. This implies condition $p$ only needs to be checked if and only if condition $q$ is true. It also yields a corresponding values of $\xi$ and $\sigma$ to be used in the last three conditions. We then check rest of the conditions based on minimum value of $f_{SSB}$.
\par We would like to assert the usefulness of these conditions. Using the above conditions not only ensures the boundedness of the potential but also gives the minimum with desired VEV alignment. The results derived here are general in nature and reduce to those obtained in \cite{Dev:2018foq} for their choice of parameters\footnote{Setting $\lambda_{2,4}$, $\rho_4$, $\alpha_i$'s and $\beta_i$'s to 0}. In \cite{Dev:2018foq}, their derived conditions for good vacuum are asserted to be only sufficient but not necessary and same holds for our case. Even with good vacuum conditions, they do not get a correct vacuum in their numerical study at all times. This possibly happens due to the parameter range of non-zero $\alpha_i$'s in their numerical analysis that leads to the violation of condition on mass-squares ${\mu}^2$ derived in this work.
\par Given the treatment here, we can also generalize the gauge-independent conditions for correct vacuum in the left-right symmetric model as:
\begin{empheq}[box=\widefbox]{gather}
\text{Tr}[ \langle \Phi \rangle\langle \Phi \rangle] \neq 0  \nonumber \\ 
\text{Tr}[\langle\Delta_{L}\rangle\langle\Delta_{L}\rangle]=\text{Tr}[\langle\Delta_{R}\rangle\langle\Delta_{R}\rangle]=0 \\ 
\text{Tr}[\langle\Delta_{L}\rangle\langle\Delta_{L}^{\dagger}\rangle]<\text{Tr}[\langle\Delta_{R}\rangle\langle\Delta_{R}^{\dagger}\rangle] \nonumber
\end{empheq}

\begin{figure}
\centering

\includegraphics[width=0.38\textwidth]{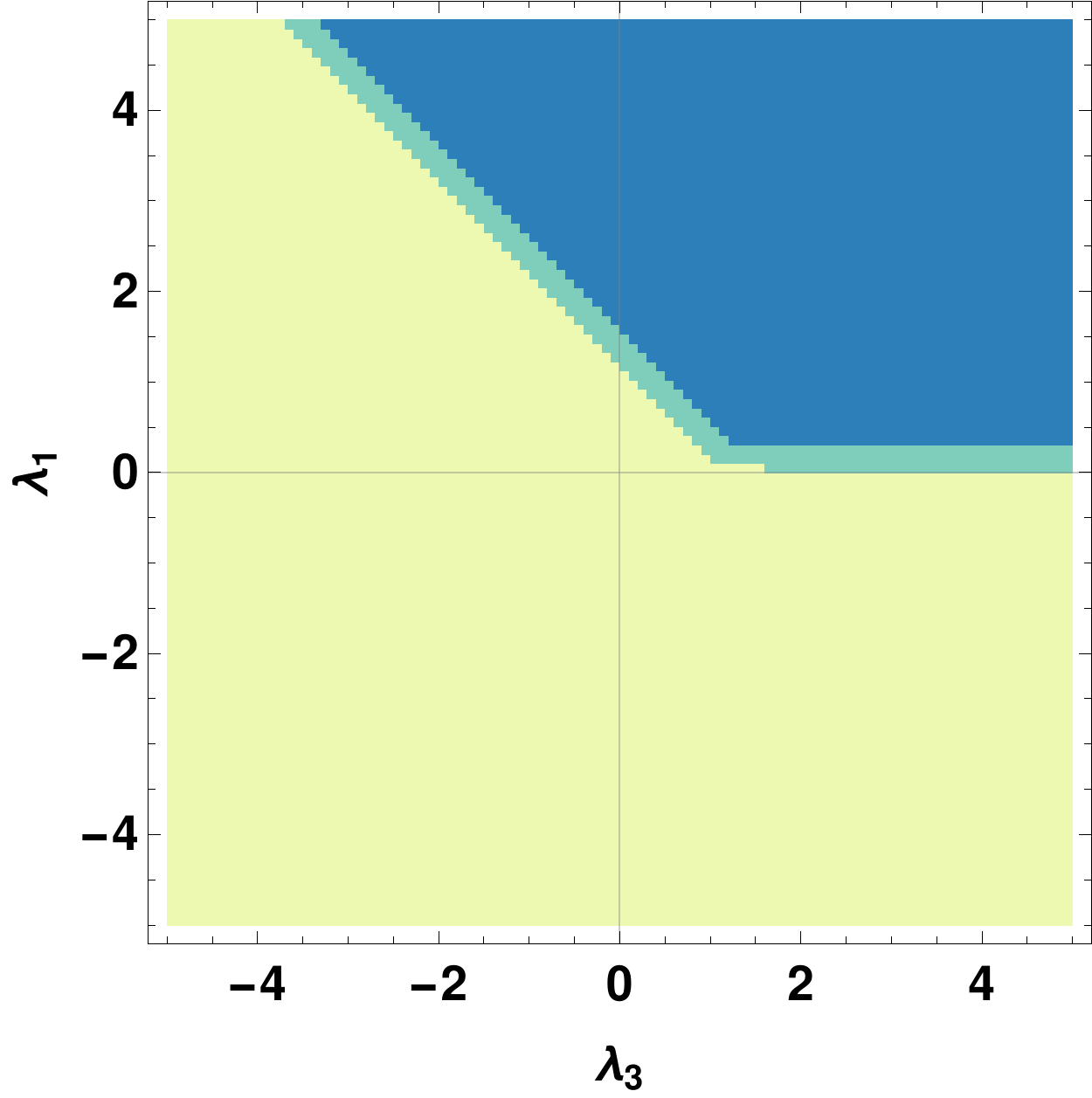}\ \includegraphics[width=0.38\textwidth]{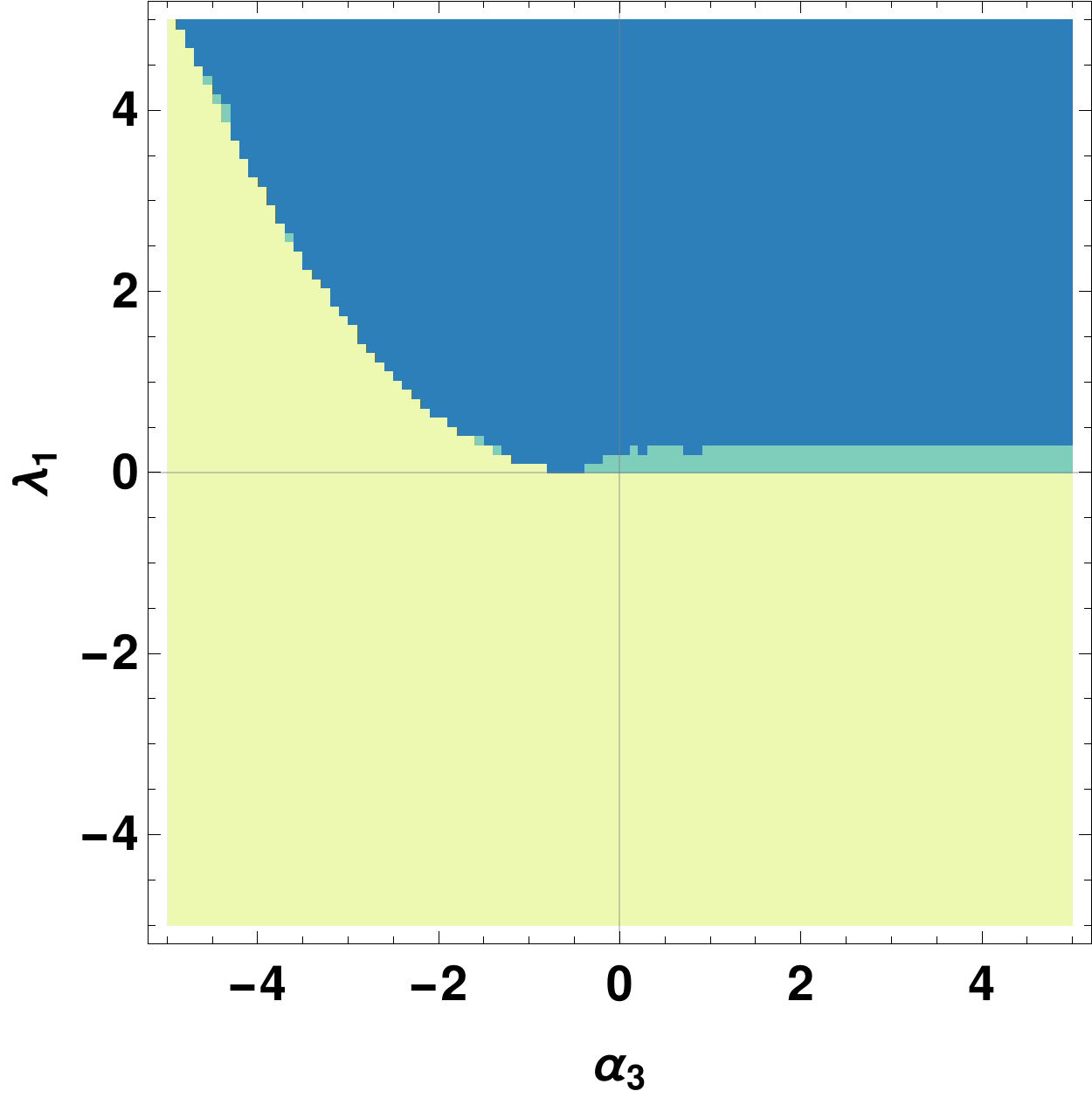}

\vspace{0.3cm}

\includegraphics[width=0.38\textwidth]{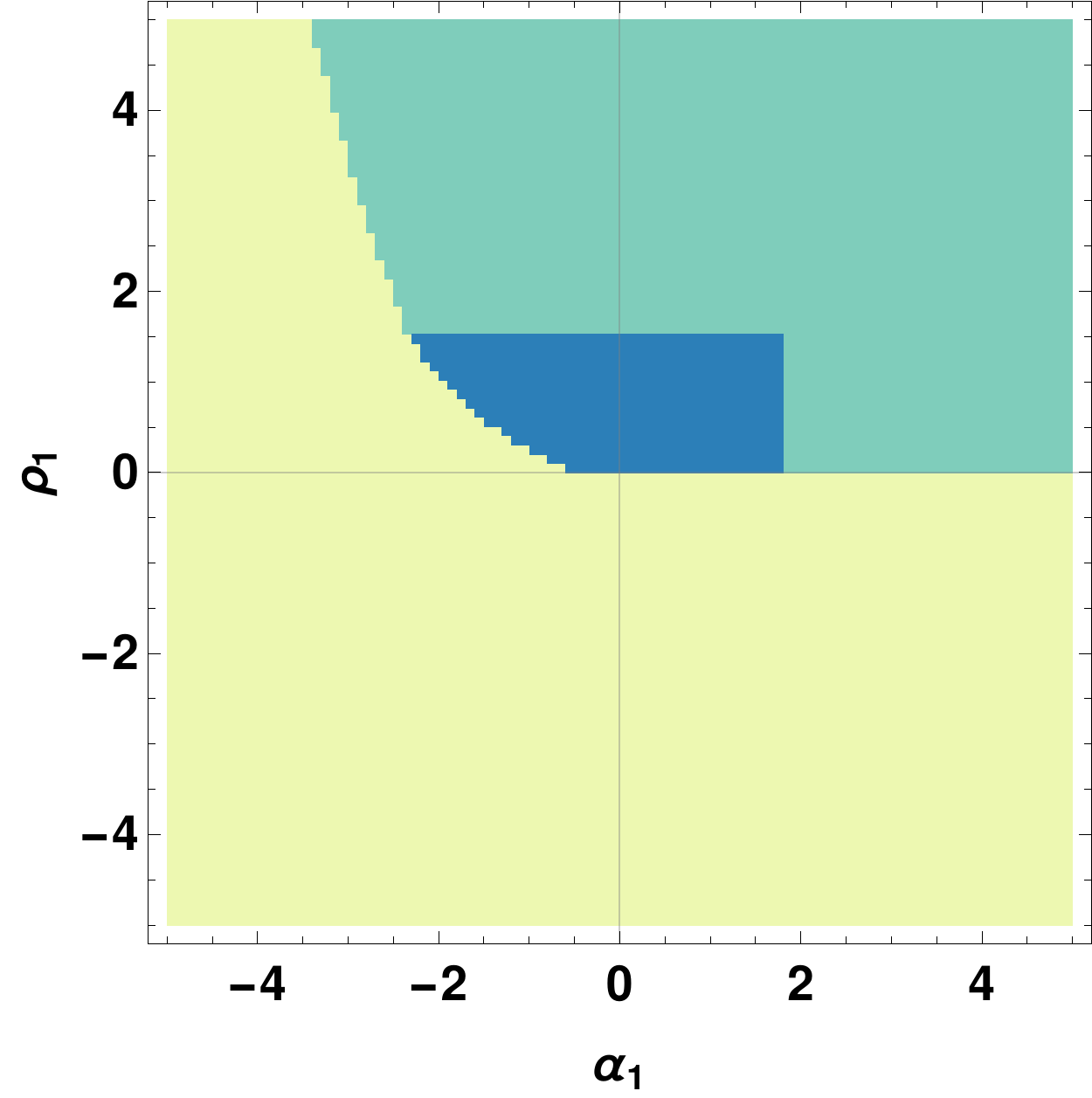}\ \includegraphics[width=0.38\textwidth]{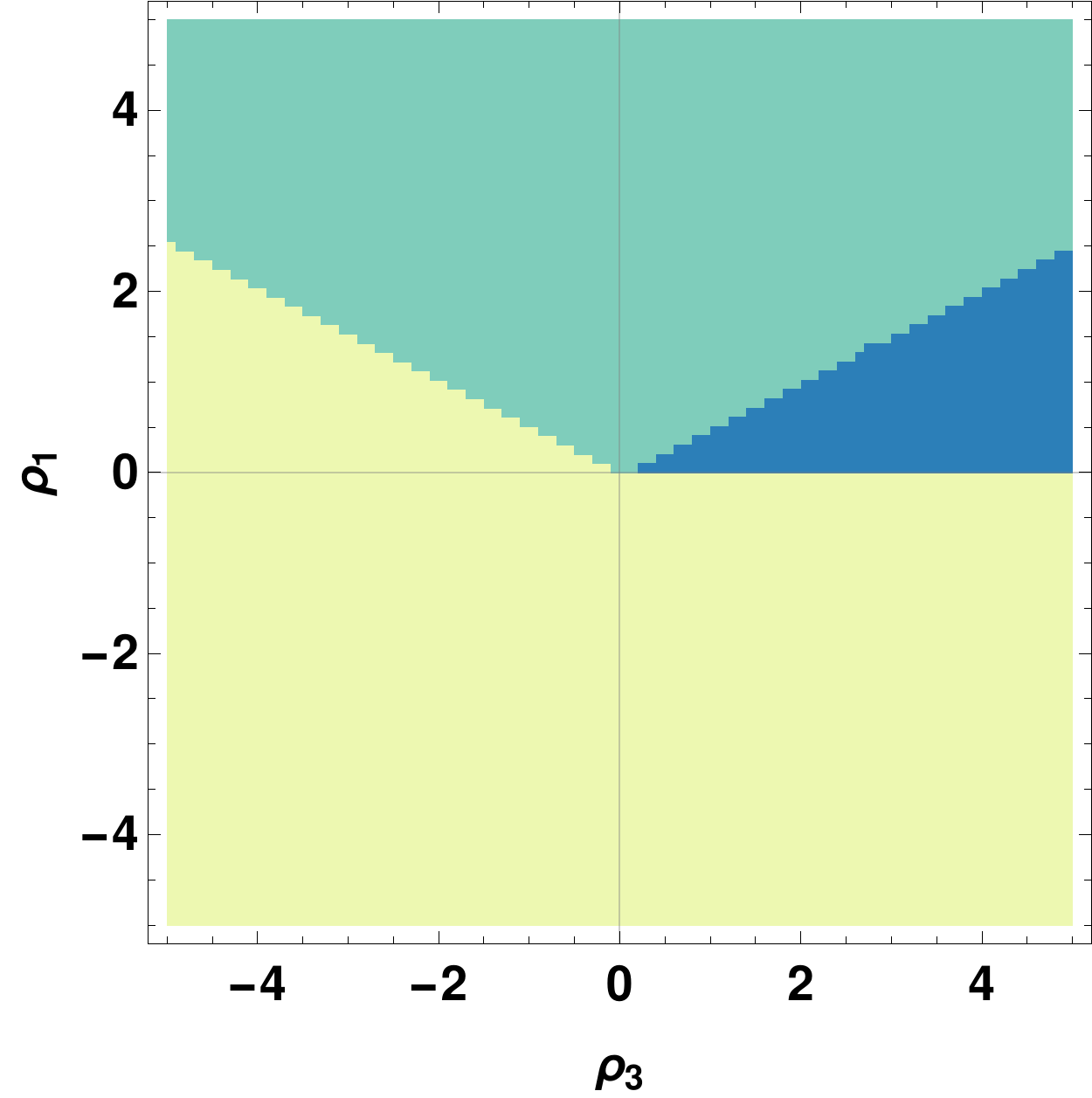}

\caption{\label{fig:MinPlots} Numerical minimization of the scalar potential of LRSM. The figures are plotted for different pair of quartic couplings  with values ranging from $(-5,5)$ and with grid pixel size of $0.1 \times 0.1$, with other quartics being set according to benchmark in Sec \ref{Num}. The yellow region indicates an unbounded potential. The green region indicates the existence of a global minimum but not with the required VEV structure. The blue region indicates the existence of a global minimum with the required VEV structure.}
\end{figure}

\section{Numerical Comparison}\label{Num}

We use the following benchmark values to study the numerical minimization of the potential and its agreement with the conditions obtained in this work.
\begin{eqnarray*}
\mu_1^2,\:\mu_2^2,\:\mu_3^2 \: & \equiv & (1,0.25,1)\, \text{TeV}^2 \\
\lambda_1,\:\lambda_2,\:\lambda_3,\:\lambda_4 & \equiv & (1,0.5,3,-0.5) \\
\rho_1,\:\rho_2,\:\rho_3,\:\rho_4 & \equiv & (1,0.5,3,-0.5)\\
\alpha_1,\:\alpha_2,\:\alpha_3 & \equiv & (0.5,0,0.5)\\
\beta_1,\: \beta_2,\:\beta_3 & \equiv & (0,0,0)
\end{eqnarray*}

In fig \ref{fig:MinPlots}, the potential is minimized for a pair of quartics with other couplings set according to the benchmark values. The minimization was performed with the \texttt{NMinimize} function with \texttt{NelderMead}, \texttt{DifferentialEvolution} and \texttt{SimulatedAnnealing} method in $\texttt{Mathematica}$. The pixel size of the grid is $0.1 \times 0.1$. With each parameter running from $(-5,5)$ yields a $50\times50$ matrix. The yellow region has unbounded minima that violates the BFB conditions. The green region is bounded and has a global minimum but with an incorrect VEV alignment. In blue region, the potential undergoes correct sponataneous symmetry breaking to the desired VEV structure of the vacuum. This vacuum is stable and is phenomenologically viable.
\par We would like to assert that the results shown in fig \ref{fig:MinPlots} are in complete agreement with the vacuum stability and correct vacuum conditions obtained in this work. It should also be noted that although conditions to exhibit SSB to correct vaccum were derived using a stronger condition, they match results from numerical minimization remarkably.

\section{Renormalization Group Equations Analysis }\label{RGE}

In a general case of randomly selected initial values, the evolution of quartic couplings according to the renormalization group equations (RGEs) for the model can lead to their running outside the allowed parameter space. Constraining the running of the quartic couplings to satisfy the vacuum stability conditions upto a certain high energy scale ensures the boundedness of the potential. In this section, we discuss some more constraints on the quartic couplings before we present an example study to demostrate the usefulness of the conditions derived earlier.

\subsection{Mass Spectrum \& Unitarity Bounds}

Along with BFB conditions and correct symmetry breaking, it's necessary to check that the potential exhibits a physical scalar mass spectrum. The scalar mass spectrum of the LRSM has 14 physical particles. It includes 8 electrically neutral \footnote{It contains two massless neutral degrees of freedom absorbed as the longitudinal polarization modes of physical gauge bosons.}, four singly-charged and four doubly-charged Higgs bosons. The scalar mass spectrum for LRSM is given below\cite{Duka:1999uc,Chakrabortty:2016wkl}:
\begin{eqnarray*}
M_{H_0^0}^2 & = & 2\left(\lambda_1-\frac{\alpha_1^2}{4\rho_1}\right) \kappa_+^2,\\
M_{H_2^{\pm}}^2 \simeq M_{A_1^0}^2 \simeq M_{H_1^0}^2 & = & \frac{1}{2} \alpha_3 v_R^2,\\
M_{H_2^0}^2 & = & 2 \rho_1 v_R^2,\\
M_{H_1^{\pm \pm}}^2 \simeq M_{H_1^{\pm}}^2 \simeq M_{A_2^0}^2=M_{H_3^0}^2 & = & \frac{1}{2} (\rho_3 -2 \rho_1) v_R^2 , \\
M_{H_2^{\pm \pm}}^2 & = & 2 \rho_2 v_R^2 + \frac{1}{2}  \alpha_3 \kappa_+^2 
\end{eqnarray*}
where $\kappa_+^2=\kappa_1^2+\kappa_2^2$. The lightest neutral scalar $M_{H_0^0}$ that only depends on the VEV of bidoublet $\Phi$ is identified as the SM Higgs boson. We have taken the best fit value of $M_{H_0^0}=m_h=125$ GeV \cite{Aad:2015zhl}. $H_1^0$, $A_1^0$ and $H_2^\pm$ are the CP-even and CP-odd neutral components and the two singly-charged scalars respectively from the bidoublet $\Phi$. $H_2^0$, $H_3^0$, $A_2^0$, $H_1^\pm$, $H_1^{\pm\pm}$ and $H_2^{\pm\pm}$ are the two CP-even and one CP-odd neutral components, 2 singly-charged and 4 doubly-charged scalars respectively from the triplets  $\Delta_L$ and $\Delta_R$. 
\par There are strong experimental bounds on most scalar masses in LRSM. This places lower bounds on the allowed values of corresponding quartic couplings in the potential as a function of the breaking scale. The heavy neutral scalars $H_1^0$, $A_1^0$ can contribute to $B_d-\overline{B}_d$, $B_s-\overline{B}_d$ and $K_0-\overline{K}_0$ mixings due to presence of tree-level FCNC couplings to the SM quarks in LRSM. Thus, there are stringent limits on their masses from the FCNC constraints \cite{Ecker:1983uh,Zhang:2007da,Maiezza:2010ic}.
$$ M_{H_1^0,A_1^0} > 15 \text{ TeV}$$
The cleanest detection channel for doubly-charged Higgs bosons is its decay to same-sign charged dilepton pairs . The current bounds on mass limits are from LHC 13 TeV run data \cite{Aaboud:2017qph,CMS-PAS-HIG-16-036}, which largely depends on charged lepton flavors involved in the decay process :
$$M_{H_1^{\pm \pm}} \gtrsim (770-870) \text{ GeV}  \qquad M_{H_2^{\pm \pm}}\gtrsim (660-760) \text{ GeV}$$
Parameter space for quartic couplings can be further squeezed by requiring tree-level unitarity to be preserved in a variety of scattering process. We consider the unitarity bounds only from 2-body scalar scattering processes \cite{Chakrabortty:2016wkl}, given below : 
\begin{eqnarray*}
&&\lambda_1 < 4 \pi/3,~ (\lambda_1+4 \lambda_2+2 \lambda_3) < 4 \pi,\\
&&(\lambda_1-4 \lambda_2+2 \lambda_3) < 4 \pi,\\ 
&&\lambda_4 < 4 \pi/3,\\
&&\alpha_1 < 8 \pi, ~\alpha_2 < 4 \pi,~ (\alpha_1+\alpha_3) < 8 \pi,\\
&&\rho_1 < 4 \pi/3, ~( \rho_1+\rho_2) < 2 \pi,~ \rho_2< 2\sqrt{2}\pi,\\
&&\rho_3 < 8\pi, ~\rho_4 < 2\sqrt{2}\pi
\end{eqnarray*}

\subsection{Example Study}\label{EX}

\begin{figure}
\centering

\includegraphics[width=\textwidth]{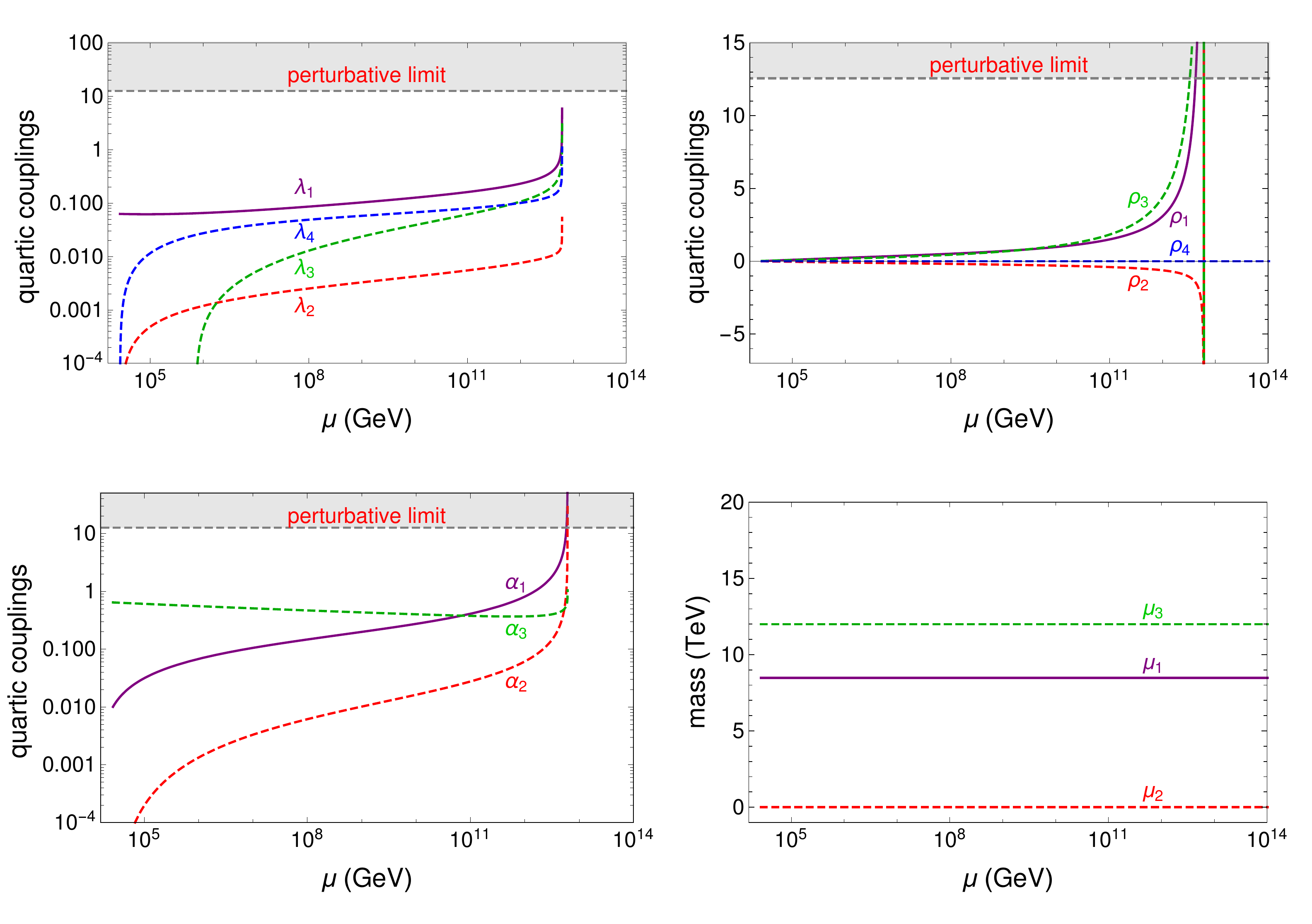}

\caption{\label{fig:RGEPlots} RG running of the quartic couplings for the benchmark in sec \ref{EX} from $v_R = 26.8$ TeV, with $r_g=\frac{g_R}{g_L}= 1.2$.}
\end{figure}

We use the following benchmark values for RGE running of the quartic couplings.
\begin{eqnarray}\label{eq:bk2}
\mu_1^2,\:\mu_2^2,\:\mu_3^2 \: & \equiv & ((8.48)^2,0,(11.99)^2)\text{ TeV}^2 \nonumber \\ \nonumber
\lambda_1,\:\lambda_2,\:\lambda_3,\:\lambda_4 & \equiv & (0.0625,0,0,0) \\
\rho_1,\:\rho_2,\:\rho_3,\:\rho_4 & \equiv & (0.01,0.0005,0.0226,0)\\ \nonumber
\alpha_1,\:\alpha_2,\:\alpha_3 & \equiv & (0.01,0,0.64)\\ \nonumber
\beta_1,\: \beta_2,\:\beta_3 & \equiv & (0,0,0)
\end{eqnarray}
The above benchmark is in complete agreement with the current experimental bounds on the scalar masses at the breaking scale. 
$$ \kappa_+=\sqrt{\kappa_1^2+\kappa_2^2}=246\text{ GeV},\:v_L=0\text{ TeV},\: v_R=26.8\text{ TeV}$$
Most importantly the ground state of the potential exhibits correct VEV structure of the theory at the right-handed breaking scale $v_R$. This is evident as the benchmark eq. \eqref{eq:bk2} satifies conditions derived for SSB to correct vaccum. 
\par We now have a complete set of initial values and the system of RGE's at one-loop level for the LRSM \cite{roth,Chakrabortty:2016wkl,Chauhan:2018uuy,Kobakhidze:2013pya}. We run the system from the breaking scale $v_R$ to the GUT scale while checking vacuum stability, perturbativity and unitarity bounds \cite{Chakrabortty:2013zja,Chauhan:2018uuy}. The results are shown in fig \ref{fig:RGEPlots}. It can be seen that quartic couplings hit the Landau pole at a scale lower than GUT scale $10^{12}$ GeV. Although the quartic couplings respects the vacuum stability conditions and unitarity bounds nearly upto the scale just before violating the perturbativity. We observe that most quartic couplings except $\rho_4$ acquire non-zero values even if set to zero at the breaking scale. $\rho_2$ is the only quartic that is observed to run to negative values although initialized at a positive value. Also notice that mass-squares $\mu^2$ don't run appreciably once set at the breaking scale. 
\par It should be mentioned that value of $r_g=\frac{g_R}{g_L}$ is also crucial to the system of RGEs. Lower values of $r_g$ for the benchmark in consideration leads to violation of vacuum stability conditions and hence an unbounded potential at high-energies. 

\section{Conclusion}\label{end}

We develop a method to extract necessary and sufficient conditions to ensure vacuum stability in LRSM by using the application of gauge orbit parameters in two-Higgs fields case. We also show application of copositivity criteria and its usefulness in simplifying the analysis for vacuum stability. 
\par As it was asserted earlier, only requiring vacuum stability does not ensure SSB to a vacuum which reproduces SM at low-energies. For this purpose, we extend the vacuum stability analysis to help yield conditions sufficient to achieve SSB to the correct vacuum which should be charge conserving and also parity violating at low-energies. These analytic techniques can also be extended to analyze metastability of the vacuum and one-loop effective potentials.   
\par We also compared our analytic results from those generated by numerical minimization of the potential. It is observed that the derived conditions are in excellent agreement with the numerical results. We also show that vacuum stability constraints along with other theoretical constraints (pertubativity, unitarity, scalar mass spectrum) coupled with RGE analysis can help us narrow down the allowed parameter space for the quartic couplings in the potential. A comprehensive study is required to explore the existence of sets of quartic and gauge couplings that obey these combined bounds. This is beyond the scope of this paper and can be another viable future direction for investigation.  

\section*{Acknowledgements}
I thank P. S. Bhupal Dev for his useful comments and Guillermo Gambini for carefully reviewing the manuscript. I also thank K. S. Babu and Xun-Jie Xu for useful discussions and Yongchao Zhang for help with fig \ref{fig:RGEPlots}. This work is supported by the US Department of Energy under Grant No. DE-SC0017987. I would also like to thank The Abdus Salam ICTP, Italy for their generous hospitality during the completion of this work.
\bibliographystyle{JHEP}
\bibliography{VS_SSB_LRSM}

\end{document}